\documentclass[letterpaper,prc,twocolumn,showpacs,floatfix,nofootinbib,preprintnumbers,superscriptaddress,amsmath,amssymb]{revtex4-2}
\usepackage{hyperref}
\usepackage{amsmath}
\usepackage{amssymb}
\usepackage{graphicx}
\usepackage[dvipsnames]{xcolor}
\usepackage{algorithm}
\usepackage{algpseudocode}
\usepackage{comment}

\newcommand{\nclus}{n_{\rm clus}}
\newcommand{\perplex}{{\tt Perp}}

\begin{document}

\preprint{LLNL-JRNL-2011023}

\title{Learning nuclear cross sections across the chart of nuclides with graph neural networks}

\author{Hongjun Choi}
\email{choi22@llnl.gov}
\affiliation{Lawrence Livermore National Laboratory, Livermore, California 94551, USA}

\author{Sinjini Mitra}
\affiliation{Arizona State University, 1151 S Forest Ave. Tempe, AZ 85281, USA}

\author{Jason Brodksy}
\affiliation{Lawrence Livermore National Laboratory, Livermore, California 94551, USA}

\author{Ruben Glatt}
\affiliation{Lawrence Livermore National Laboratory, Livermore, California 94551, USA}

\author{Erika Holmbeck}
\affiliation{Lawrence Livermore National Laboratory, Livermore, California 94551, USA}

\author{Shusen Liu}
\affiliation{Lawrence Livermore National Laboratory, Livermore, California 94551, USA}

\author{Nicolas Schunck}
\affiliation{Lawrence Livermore National Laboratory, Livermore, California 94551, USA}

\author{Andre Sieverding}
\affiliation{Lawrence Livermore National Laboratory, Livermore, California 94551, USA}

\author{Kyle Wendt}
\affiliation{Lawrence Livermore National Laboratory, Livermore, California 94551, USA}

\begin{abstract}
In this work, we explore the use of deep learning techniques to learn how 
nuclear cross sections change as we add or remove protons and neutrons. As a 
proof of principle, we focus on the neutron-induced reactions in the fast 
energy regime. Our approach follows a two-stage learning framework. First, we apply representation learning to encode cross section data into a latent space using either variational autoencoders (VAEs) or implicit neural representations (INRs). Then, we train graph neural networks (GNNs) on the resulting embeddings to predict missing values across the nuclear chart by leveraging the topological structure of neighboring isotopes.
We demonstrate accurate cross section predictions within a $9 \times 9$ block of missing nuclei. We also find that the optimal GNN training strategy depends on the type of latent representation used, with VAE embeddings performing best under end-to-end optimization in the original space, while INR embeddings achieve better results when the GNN is trained only in the latent space.
Furthermore, using clustering algorithms, we map
groups of latent vectors into regions of the nuclear chart and show that VAEs 
and INRs can discover some of the neutron magic numbers. These findings suggest 
that deep-learning models based on the representation encoding of cross sections 
combined with graph neural networks holds significant potential in augmenting 
nuclear theory models, e.g., by providing reliable estimates of covariances of 
cross sections, including cross-material covariances.
\end{abstract}

\maketitle

\section{Introduction}

Nuclear cross sections encode the probability of a subatomic particle to 
interact with an atomic nucleus, defined by its proton and neutron numbers 
$(Z,N)$ \cite{blatt1979theoretical}. Each atomic nucleus can be involved in 
many different types of nuclear reactions often simply denoted as 
{\em (incoming process, outgoing process)}, also called ``channels''. For 
example, the $(n,\gamma)$ channel involves the absorption of a neutron and the 
emission of one or several photons ($\gamma$ rays) while the $(n,f)$ channel 
involves the absorption of a neutron resulting in the fission of the nucleus.
Nuclear reaction libraries are formed by the set of all known cross sections 
for all known nuclei \cite{iwamoto2023japanese,plompen2020joint,brown2018endf}. 
Such libraries are essential inputs to nuclear technology applications 
\cite{bernstein2022first}, which include energy production 
\cite{bolstelmann2022nuclear}, nuclear medicine 
\cite{khandaker2022significance}, and nuclear forensics 
\cite{hayes2017applications}. They also play a critical role in our attempts at 
understanding the formation of elements in the universe 
\cite{roederer2023element,mumpower2016impact}. The importance of such 
libraries, and of nuclear data in general, in scientific research was recently 
underlined in the 2023 Long Range Plan for Nuclear Science commissioned by the 
U.S.\ Department of Energy and the National Science Foundation 
\cite{dodge2023lrp}.

Measuring nuclear cross sections requires complex and expensive experimental 
facilities and is often impossible for short-lived radioactive nuclei that 
cannot be made into targets or created in radioactive beams. While nuclear reaction theory has greatly improved 
over the past decades \cite{thompson2009nuclear}, its level of sophistication 
varies greatly depending on the nucleus and type of reaction. In very light 
nuclei, \textit{ab initio} theory can successfully describe reactions at low 
energy in a fully quantum-mechanical framework \cite{navratil2020initio}; in 
contrast, reactions of heavy nuclei with neutrons with just a few sub-eV energy 
already lead to the formation of a highly excited nucleus with an extremely 
complex structure that can only be described within a statistical phenomenology framework 
\cite{hodgson1987compound,carlson2014theoretical}. In practice, such 
statistical reaction theory relies on additional theoretical inputs describing 
nuclear properties such as the nuclear level density; the probability of 
capturing incident particles; the rate of decay through the emission of 
photons, neutrons, protons; etc.\ \cite{capote2009ripl}. Today the best nuclear 
data libraries on the market largely rely on phenomenological models to 
determine such properties. These models are carefully calibrated to match 
realistic application requirements such as the criticality regimes of a nuclear 
reactor, at the price of a loss in generality and predictive power of the 
underlying theory \cite{goriely2019reference,brown2018endf}. As a result, 
current models fail at consistently predicting the complex relationships 
between different types of cross sections in a single nucleus, or between cross 
sections in different nuclei across the chart of isotopes. This lack of 
predictive power is a significant impediment to efforts at designing new, 
cleaner, or more efficient nuclear technologies. While research in fundamental 
reaction theory continues to progress, techniques originating from artificial 
intelligence and machine learning research offer an alternative viewpoint 
\cite{kolos2022current, schwartz2103modern, feickert2102living, 
fox2024illuminatingsystematictrendsnuclear}.

Indeed, each single cross section is a continuous one-dimensional function of 
the energy of the incident particle (akin to a time sequence) labeled by the 
number of protons $Z$ and neutrons $N$. In neutron-induced reactions in the 
fast regime---when the energy $E$ of the incident neutron is $E>100$ keV---
cross sections are rather smooth functions of neutron energy (notwithstanding 
threshold effects). Therefore, dimensionality reduction techniques could in 
principle be applied to compress information and learn from the data the 
physically relevant latent space that underpin the cross sections.

In addition, cross sections of nuclei with close $(Z,N)$ numbers often tend to 
have similar forms, suggesting a correlation with the position on the 2D grid 
structure spanned by $Z$ and $N$. This observation suggests employing geometric 
deep learning for learning meaningful representations in the latent space, a 
technique that has recently emerged as an impactful sub-field adept at learning 
non-Euclidean sub-spaces~\cite{bronstein2017geometric}. Graph Neural Networks 
(GNNs) \cite{scarselli2008graph} in particular provide an easy way to perform 
node-level, edge-level, and graph-level prediction tasks while being 
permutation-invariant for structured grid-like data in such non-Euclidean 
sub-spaces, as has been shown for particle reconstruction \cite{ju2020graph}.

The goal of this paper is to 
effectively learn the intricate relationships between cross 
sections of different nuclei by leveraging such state-of-the-art deep learning 
models. To this end, we employ two complementary techniques in succession. As 
the first step, we use either a variational autoencoder (VAE) 
\cite{doersch2016tutorial} or an implicit neural representation (INR) to 
uncover the latent space of nuclear cross sections and reduce the 
dimensionality of the problem. 
In the second step, we use GNNs to account for the implicit 2D geometric nature of the nuclear chart, where the graph neighborhood is defined by similar $Z$ and $N$ numbers, and the learned latent space are used as graph node features.
The resulting network provides an 
efficient and consistent framework to predict unknown cross sections not too far 
from the training region.

This paper is organized as follows: Section~\ref{sec:theory} gives a broad 
overview of the challenges facing nuclear theory when computing nuclear data.
In Section~\ref{sec:methods}, we summarize the main aspects of our theoretical 
approach, including the architecture of the network and the characteristics of 
the datasets used for training. Section~\ref{sec:results} shows the results for 
different GNN architectures and encoder approaches. 
Section~\ref{sec:discussion} discusses our results in the context of related 
work on nuclear data before we summarize our findings in 
Section~\ref{sec:conclusion}.


\section{Theoretical models of nuclear data: Challenges}
\label{sec:theory}

As mentioned in the introduction, nuclear reaction libraries are formed by the 
set of all known cross sections for all known nuclei. Most of this data is not 
simply a set of experimental data points but the result of an {\em evaluation} 
of all available data by subject matter experts. This evaluation is itself the 
outcome of a complex, multi-stage, iterative process that involves: 
(i) carefully analyzing available experimental results (vetting outliers, 
estimating uncertainties, etc.); (ii) adjusting model parameters of nuclear 
reaction codes such as TALYS \cite{koning2023talys} or EMPIRE 
\cite{herman2007empire} in such a way that the cross sections computed with 
such codes ``match'', or are close to, experimental measurements; (iii) 
{\em processing} the data (energy binning, heating of cross sections to room 
temperature values, etc.); and (iv) verifying the nuclear reaction library through 
{\em benchmarks}, e.g., running a suite of codes simulating critical assemblies.

From a physics perspective, nuclear reactions are extremely complex processes. 
Ideally, one would like to describe any reaction as a quantum-mechanical 
evolution from an initial state, e.g., an incoming particle $a$ with energy 
$E_a$, spin $J_a$, and parity $\pi_a$ impinging on a target at rest $A$ with 
energy $E_A$, spin $J_A$, and parity $\pi_A$ to a final state such as two 
reaction products $b$ and $B$ with respective energies $E_b$ ($E_B$), spins 
$J_b$ ($J_B$), and parities $\pi_b$ ($\pi_B$). Such a description would require the 
perfect characterization of the quantum-mechanical state of all reaction 
products $a$, $A$, $b$, and $B$ as well as the exact knowledge of the nuclear 
many-body Hamiltonian $\hat{H}$, since the unitary quantum-mechanical time 
evolution operator is $e^{-i\hat{H}t/\hbar}$. Both are out of reach of current 
nuclear theory.

In practice, the physics of a nuclear reaction is instead described by a suite 
of models with varying degrees of sophistication and phenomenology. These 
models play an essential role in the evaluation process and can be divided into 
several broad categories: (i) optical models describe the interaction of an 
incoming particle $a$ with a target $A$; (ii) level density models characterize 
the number of available energy levels $\rho(E,J,\pi)$ at a given excitation 
energy $E$, angular momentum $J$, and parity $\pi$; (iii) gamma-strength 
functions $\gamma_{SF}(E)$ characterize the probabilities of emitting photons 
as a function of energy $E$; (iv) fission transmission coefficients quantify the 
probability for a (heavy) nucleus to undergo fission as a function of energy, 
angular momentum, and parity; and (v) pre-equilibrium reaction models describe  
interaction mechanisms between projectile and target that involve specific 
nuclear excited states. 

All these models are but different facets of the ideal quantum-mechanical 
evolution described above. This implies that they should all be related to one 
another, at the very least indirectly. For example, let us assume one were to 
describe the structure of an atomic nucleus with a self-consistent mean-field 
model based on some effective nucleon-nucleon potential 
\cite{bender2003selfconsistent}. How this mean field is computed or parametrized 
should directly impact the optical potential (the real part of the optical 
potential does represent the nuclear mean field), the nuclear level density 
(through the excited states associated with the mean-field potential), the 
gamma-strength-function (which is typically related to functional derivatives 
of the mean field with respect to nuclear densities), fission transmission 
coefficients (which depend on how the mean-field potential changes with 
deformation), etc. Moreover, since the same effective nucleon-nucleon potential 
would be used to compute every single nucleus, nuclear properties would also be 
heavily correlated as a function of proton and neutron number. 

Therefore, even if such a theoretical model still contains parameters fitted 
to experimental data---in this case, the parameters of the effective 
nucleon-nucleon force---all nuclear properties are directly related to that 
single set of parameters. In a sensitivity analysis, one would find that 
variations of the model parameters result in {\em correlated} variations of 
the model outputs, i.e., the nuclear data properties of interest. In current 
approaches based on separate models for each type of nuclear data, such 
correlated variations, which manifest relationships between different data, 
are impossible to capture with nuclear theory alone.

Instead, each nuclear model is treated as a separate black box independently 
parametrized by a handful of parameters. In a typical evaluation process, 
these parameters are adjusted in such a way that nuclear cross sections match 
experimental measurements and the resulting benchmarks are satisfied 
\cite{plompen2020joint,brown2018endf,koning2019tendl,iwamoto2023japanese}. 
These fits will often take the form of a functional dependency of model 
parameters in terms of proton and neutron number (or atomic mass number $A$), 
excitation energy, or angular momentum. The nuclear reaction codes mentioned 
above typically implement a few different variants of such systematics. By 
construction, these systematics become highly questionable in nuclei or at 
energies that are beyond the range of their fit, which directly impacts the 
reliability of nuclear data libraries for extrapolations. 

By employing a collection of independent nuclear models locally 
calibrated to experimental data, it thus becomes nearly impossible to (1) make 
reliable predictions outside the range of currently available measurements and 
(2) capture the relationships (or correlations) between nuclear data within 
the same nucleus, or between different nuclei. In this paper, we will show 
that deep-learning techniques provide an alternative solution to both points.

\begin{figure*}[!htbp]
\centering
\includegraphics[width=0.85\linewidth]{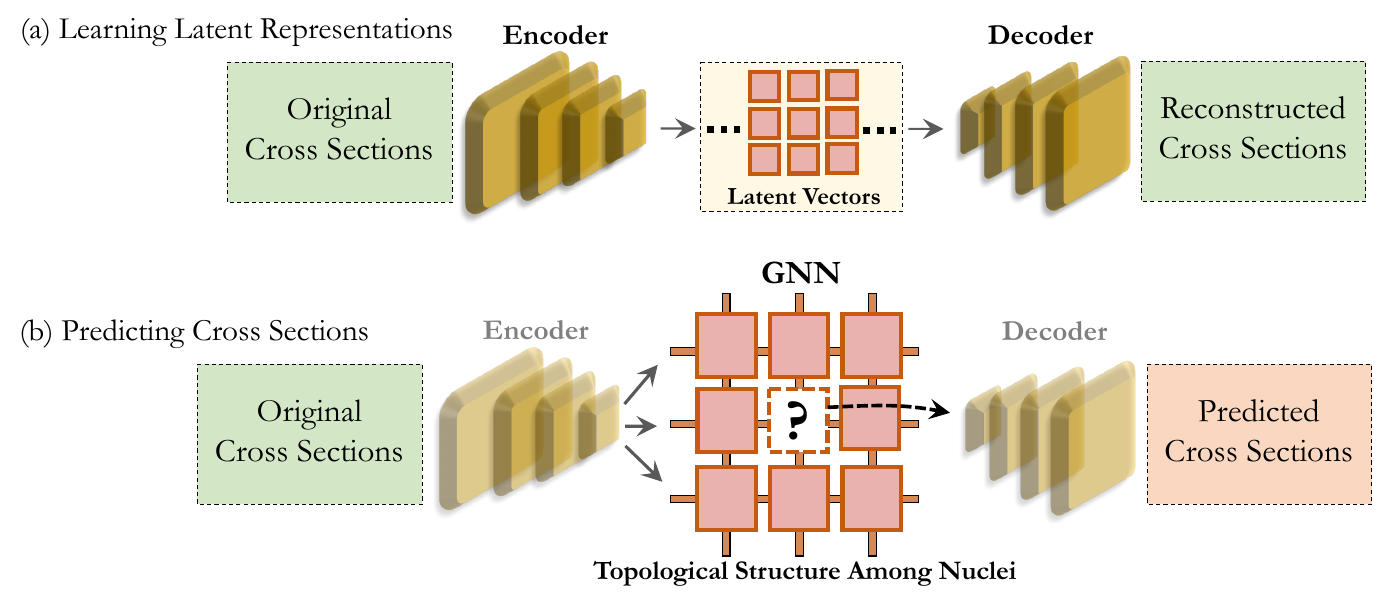}
\caption{Schematic representation of our framework. The initial data is a (set 
         of) nuclear cross sections represented as a vector indexed by the 
         energy of the incident particle. Panel (a): The data is encoded into a latent 
         representation using either a variational autoencoder (VAE) or an 
         implicit neural representation (INR) network (representation learning).
         Panel (b): A graph neural network (GNN) then learns how the data in the latent 
         space in nucleus $(Z,N)$ transforms into neighboring nuclei in order 
         to make predictions in nucleus $(Z',N')$. The decoder then transforms 
         the predicted vector in the latent space back into the space of 
         nuclear cross sections (latent space predictions).} 
\label{fig:overview}
\end{figure*}


\section{Methodology}
\label{sec:methods}

The limited number of available data for individual cross sections together 
with the potentially complex relationships between cross sections of different 
nuclei can make the training of deep learning models challenging. To overcome 
such a challenge, we separate the overall learning problem into two 
interconnected components, namely a representation learning task and a 
prediction task, as shown in Fig.~\ref{fig:overview}. In the representation 
learning task, we are learning the low-dimensional, latent space representation 
of cross sections; in the prediction task, we make predictions to unknown 
nuclei from such latent representations and use the decoder trained in the 
first task to bring them back into the original space of cross sections.
In contrast to the space of cross sections, the latent space provides a more 
compact yet meaningful representation that allows the model to learn more 
rapidly and generalize better. Additionally, the separation of these two steps 
ensures a clear delineation between the process of learning the representation, 
which may require data augmentation, i.e., interpolation and scaling of the 
cross section data, and the actual prediction tasks, which should only involve 
the cross section data that physically exist, i.e., a cross section that is not 
augmented with additional, synthetic data or modified.

As schematically illustrated in Fig.~\ref{fig:overview}(a), the representation 
learning component will focus on learning the underlying latent representation 
space of all known cross sections. In the following discussion, we note 
$\sigma_{\alpha}(E)$ the cross section of the nuclear reaction, or channel, 
$\alpha$ as a function of the energy $E$ of the incident particle, where the 
energy is discretized on a finite grid $E \rightarrow \{ E_{o} \}_{o=1,m}$, 
where $m$ is the dimension of the original data vector. For the encoding of the 
cross section, we train either a 1D-convolutional VAE \cite{kingma2013auto} or 
an encoding-decoding hypernetwork \cite{ha2016hypernetworks} that predicts the 
weights for the INR model \cite{sitzmann2020implicit} on data from $s$ nuclei 
to learn a $k$-dimensional latent space for various cross sections. By design, 
$k < m$ so that the size of the feature matrix is reduced to $ \mathbf{X} \in 
\mathbb{R} ^{s \times k}$ while keeping the implicit grid intact.

For the prediction task shown in Fig.~\ref{fig:overview}(b), we rely on a GNN 
approach. The available cross section data in the latent space can be 
represented as a graph $\mathcal{G}=(\mathcal{V, E})$, where each nucleus is 
represented by a node $v_i \in  \mathcal{V}$ (the set of total nodes) while the 
relationship between nodes $v_{i},v_{j}$ constitutes an edge $e_{ij} \in 
\mathcal{E}$ (the set of edges). Each node $v_i$ with $i\in[0,s-1]$ consists of 
a $k$-dimensional feature vector $\mathbf{x}_i$, defining the feature matrix 
$\mathbf{X}$ $ \in \mathbb{R} ^{s \times k}$.


\subsection{Dataset}
\label{subsec:data}

Our dataset is a subset of the TENDL dataset of nuclear cross sections 
\cite{koning2019tendl}. In practice, we only consider the $(n,el)$ (i.e., 
elastic), $(n,n')$ (i.e., inelastic), $(n,\gamma)$, and $(n,2n)$ cross sections 
for all nuclei included in TENDL. Since cross sections are probabilities per 
unit area, their values are always positive: $\sigma_{\alpha}(E) \geq 0$ for 
all channels $\alpha$. The TENDL dataset used contains $s=2171$ nuclei and can be 
represented in a 2D grid corresponding to the chart of isotopes and labeled as 
$(Z,N)$ with $Z$ the proton number and $N$ the neutron number of the nucleus.
For the elastic, inelastic, $(n, \gamma)$, and $(n,2n)$ channels the original 
cross section is defined on an energy grid of $m = 256$ points regularly spaced 
between 0 and 20 MeV. The values of each type of cross section are normalized 
by dividing by their maximum across the dataset prior to network training. After training, the data 
is then embedded into a lower $k$-dimensional latent space with $k=32$ using 
either a pre-trained VAE or INR model, resulting in a final data matrix of size 
$(s\times k) = (2171\times 32)$, which serve as input for the downstream prediction task. 

The size of the TENDL dataset is rather small to train deep neural networks. This can be mitigated using data augmentation techniques.
In the representation learning task, we only focus on the cross section data, 
i.e., describing curves, rather than any underlying physical relationship and 
properties. The goal is simply to construct a continuous, multi-dimensional 
latent space of curve shapes. Since we do not directly assign any physical 
properties to such a space, we can adopt an interpolation-based data 
augmentation strategy, in which we produce new training examples by linearly 
interpolating between two randomly picked cross section samples with weights 
that add up to 1.0. Essentially the latent space thus generated provides a 
super-set of all valid cross sections while at the same time allowing for a 
smooth transition between them for better optimization in the cross section 
prediction task with the GNN. For the latter, only actual cross sections in 
existing nuclei are used as training data. We found that this type of data augmentation improves the encoding performance of the VAE model, whereas the INR model does not benefit from it during training.


\subsection{Representation Learning}
\label{subsec:latent}

Representation learning aims to condense raw input data into a concise format that 
preserves crucial information. Ideally, the learned 
representation provides a lower-dimensional space which makes subsequent 
learning tasks more efficient. In our setup, the goal of representation 
learning is to learn the latent space of all cross sections. By doing so, not only can 
all known cross sections of a nucleus be encoded in the latent space, 
we can also predict new cross sections for known and unknown nuclei.
This prediction task then just translates into the task of finding the right 
vector in the learned representation space.

\begin{figure*}[t]
\centering
\includegraphics[width=0.80\linewidth]{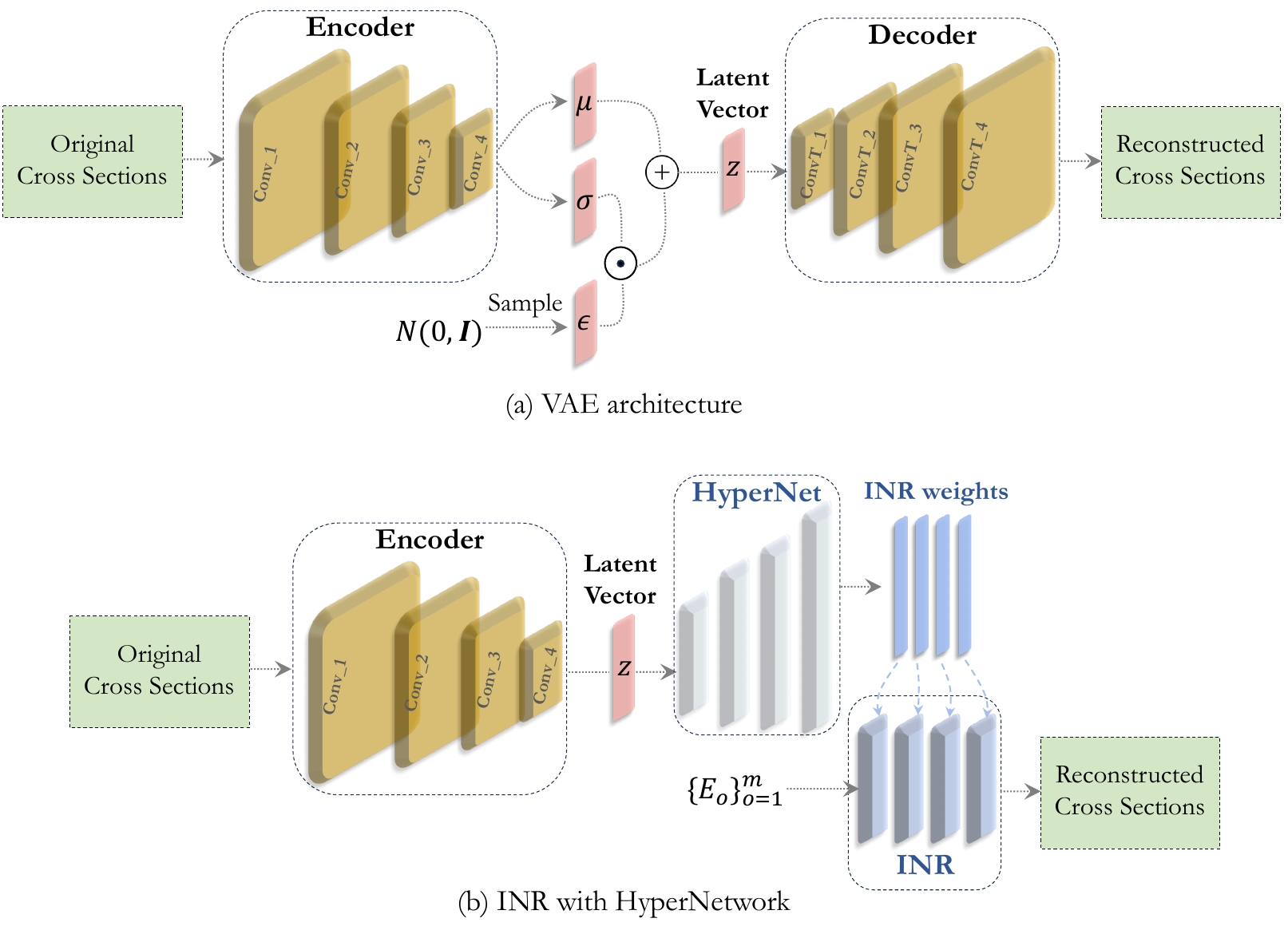}
\caption{Panel (a): Variational Autoencoder framework illustrating the encoding and 
         decoding process with a latent space bottleneck. Panel (b): Implicit Neural 
         Representation (INR) architecture integrated with a hypernetwork 
         (HyperNet). The hypernetwork generates weights for the INR, informed 
         by a convolutional encoder that processes input data. Note that the 
         INR takes input coordinates, $\{e_{i}\}$ (representing the energy index with a 
         length of $m=256$) and maps them to the corresponding cross sections.}
\label{fig:vae_inr}
\end{figure*}


\subsubsection{Variational Autoencoder}
\label{subsubsec:vae}

One of the most commonly used approaches for representation learning is an 
autoencoder (AE) architecture. An AE is a neural network designed to encode 
input data into a lower-dimensional space (i.e., bottleneck) and then decode it 
back to the original data space. The autoencoding objective is centered around minimizing the discrepancy 
between the original input and the reconstructed output. For variational 
autoencoders (VAEs)~\cite{doersch2016tutorial}, the standard autoencoding 
objective is augmented by the inclusion of variational inference 
\cite{jordan1999introduction}, where the encoding process maps input data not 
to a fixed point but to a probability distribution in the latent space. This 
dual training objective involves both minimizing the reconstruction error and 
aligning the latent space distribution with a specified form, i.e., a standard 
Gaussian distribution (with parameters $\mu$ and $\sigma$), as illustrated in Fig.~\ref{fig:vae_inr}(a). 
Specifically, the VAE training loss can be defined by two key terms: the 
reconstruction loss and the Kullback-Leibler (KL) divergence,
\begin{multline}
\mathcal{L}_{\rm VAE} = -\mathbb{E}_{q_\phi(z|x)}[\log p_\theta (x|z)] \\ 
+ \lambda \cdot D_{\rm KL}(q_\phi(z|x) || p(z)) 
\end{multline}
The first term encourages the VAE to accurately reconstruct the original input 
data. It measures the negative log-likelihood of the decoder distribution 
($p_\theta(x|z)$) generating data that resembles the input ($x$) given a 
sampled latent code ($z$)  from the encoder distribution ($q_\phi(z|x)$). The 
second term acts as a regularizer, forcing the learned latent distribution 
($q_\phi(z|x)$) to be similar to a prior distribution ($p(z)$), i.e., a 
standard normal Gaussian. The hyperparameter $\lambda$ controls the balance 
between reconstruction accuracy and the regularity of the latent space. In this work, we set $\lambda$ to $0.1$.

The integration of variational inference ensures a continuous representation 
space and makes the network generative: we can generate diverse and realistic 
new data simply by sampling from the learned latent space distribution and 
applying the decoder. Overall, VAEs provide a powerful framework for 
unsupervised representation learning, while at the same time facilitating 
high-quality reconstruction and generation.

In this work, the encoder of the VAE model comprises a sequence of 
convolutional layers followed by Rectified Linear Unit (ReLU) activation functions 
\cite{agarap2018deep} and a flattening operation. This architecture is designed 
to progressively reduce the spatial dimensions of the inputs while increasing 
the number of feature channels. Our encoder includes four convolutional layers 
---noted conv\_1, conv\_2, conv\_3, and conv\_4 in Fig.~\ref{fig:vae_inr}(a)---
with increasing filters (32, 64, 128, 256), kernel size 4, and stride 2. We then convert the output of 
convolutional layers into a flattened vector of size $k$. 

The decoder comprises a sequence of transposed convolutional layers aimed at 
reconstructing the original inputs from the latent space. It also employs ReLU 
activations and concludes with a sigmoid activation for the reconstructed 
inputs. As the mirror opposite of the encoder, the decoder progressively 
refines the representation through four layers with specific filter sizes and 
strides: 128 filters with a kernel size of 5 and stride of 2, followed by 64 
filters with a kernel size of 5 and stride of 2, then 32 filters with a kernel 
size of 6 and stride of 2, and finally, the output layer with a kernel size of 
24 and a stride of 8. The output layer returns a vector of size $m$. By jointly 
optimizing the encoder and decoder, the VAE learns a compact 
representation of given inputs.


\subsubsection{Implicit Neural Representation}
\label{subsubsec:inr}

INR is a method for representing and modeling input data without explicitly 
defining the geometry or structure of objects \cite{mildenhall2021nerf, 
sitzmann2020implicit}. Traditional data encoding methods rely on discrete 
representations, such as data grids, which are constrained by their spatial 
resolution and are susceptible to inherent discretization artifacts. In 
contrast, INR represents input data (including sound, time-series, images, and 
3D scenes) as a continuous function of its coordinate locations, where each 
value is generated independently. This function is approximated by a deep 
neural network. INR offers flexibility and adaptability for encoding data 
independently of resolution through the use of coordinate grids, making them 
highly effective with numerous potential applications 
\cite{mescheder2019occupancy, mildenhall2021nerf, szatkowski2023hypernetworks}.

While INRs offer advantages in practical scenarios, they face the limitation of 
often struggling to generalize beyond the single input on which they were 
trained. This can hinder their effectiveness in deep learning applications, 
where robust generalization across diverse training datasets is crucial. 
To address this issue, hypernetworks have been introduced 
\cite{ha2016hypernetworks}. Hypernetworks act as meta neural networks, 
generating weights for a primary neural network, the desired INR. Sitzmann et 
al.\ \cite{sitzmann2020implicit} successfully employed a convolutional encoder 
and hypernetwork combination to guide the INR in learning more generalizable 
representations of training data. This unified framework helps the INR overcome 
the `single-fit' limitation by learning priors over the input space, enhancing 
its representational capabilities while maintaining its accurate reconstruction 
abilities.

As shown in Fig.~\ref{fig:vae_inr}(b), our framework integrates a hypernetwork 
with an INR. The convolutional encoder leverages the same configuration as in 
Fig.~\ref{fig:vae_inr}(a), featuring a series of convolutional layers. It 
processes cross section data and compresses it into a 32-dimensional latent 
vector. This vector is then passed to the hypernetwork decoder, which consists 
of standard multi-layer perceptron layers with ReLU activations. The 
hypernetwork's output provides the weights for the INR. Finally, the INR takes 
coordinates as input and reconstructs the original input data. This pipeline 
enables networks to be trained in an end-to-end manner with the 
following loss function:
\begin{equation}
\mathcal{L}_{\rm INR}=\mathcal{L}_{\rm recon}+\alpha\mathcal{L}_{\rm latent}+\beta\mathcal{L}_{\rm weights}, \label{eq:nern}
\end{equation}
where $\mathcal{L}_{\rm recon}$, $\mathcal{L}_{\rm latent}$, and 
$\mathcal{L}_{\rm weights}$ represent the reconstruction loss, regularization 
terms for latent vectors, and weights in the hypernetwork, respectively. The hyperparameters 
$\alpha$ and $\beta$ are utilized to balance the contributions of the different 
losses. Specifically, each loss term is computed as follows:
\begin{subequations}
\begin{align}
\mathcal{L}_{\rm recon}   & =\frac{1}{N_{\rm batch}}\sum_{i=1}^{N_{\rm batch}}\|\mathbf{y}_{i}-\hat{\mathbf{x}}_{i}\|_{2}, \\
\mathcal{L}_{\rm latent}  & =\frac{1}{N_{\rm batch}}\sum_{i=1}^{N_{\rm batch}}\|\mathbf{z}_{i}\|_{2}^2, \\
\mathcal{L}_{\rm weights} & =\frac{1}{N_{\rm batch}}\frac{1}{N_{\mathcal{W}}}\sum_{i=1}^{N_{\rm batch}}\sum_{\mathbf{w}\in\mathcal{W}}\|\mathbf{w}_{i}\|_{2}^2\, ,
\label{eq:loss_INR}
\end{align}
\end{subequations}
where $\| \dots \|_{2}$ is the usual Euclidean norm. Here, $\mathbf{y}_{i}$ 
represents the $i$-th ground truth input and $\hat{\mathbf{x}}_{i}$ its 
corresponding reconstruction. $N_{\rm batch}$ represents the batch size, 
indicating the number of samples processed during each training iteration.
$\mathbf{z}_{i}$ refers to the latent vector with $k$ dimensions produced by 
the encoder for the $i$-th sample in the batch. $N_{\mathcal{W}}$ represents 
the total number of parameters in the INR model, and $\mathbf{w}_{i}$ are the 
individual weights. Note that $\mathcal{L}_{\rm latent}$ and 
$\mathcal{L}_{\rm weights}$ are regularization terms that aim to decrease the 
overall magnitude of feature and weight representations, making them less prone 
to overfitting by penalizing large magnitude. Through empirical evaluation, we 
determined that setting $\alpha=0.01$ and $\beta=1.0$ results in optimal 
reconstruction performance.


\subsection{Graph Neural Networks}
\label{subsec:gnn}

GNNs are optimizable transforms that operate on all attributes of a graph 
(nodes, edges, global context, inherent relationships, etc.) while preserving 
symmetries and the underlying structure of the data. 
Hence, GNNs are well suited to expose and explore the relationships 
between nuclear cross sections across a two-dimensional grid---the nuclear chart. 
GNNs accept a 
graph as input (for example, the encoded 2D TENDL grid) with the information 
contained in the nodes (feature vectors) and perform transformations without 
changing the connectivity of the input. GNNs also produce a graph as the 
output. 

In this study, we are interested in graph-based imputation, where the goal is to predict missing node features by leveraging the structural and feature information from neighboring nodes. Two primary learning paradigms are commonly considered, transductive and inductive settings. In the transductive setting, the model is trained and evaluated on the same graph, where the node and edge structure remain fixed. In the inductive setting  a subset of the graph is used for training while testing is performed on a disjoint subset or entirely new graph. Such a paradigm generally requires much stronger generalization power of the network. In this initial work, we focus on the transductive setting, as our primary interest lies in improving predictions by fully exploiting structural correlations across neighboring nuclei within the given nuclear chart.


\subsubsection{Graph Connectivity and Message-Passing}
\label{subsubsec:arch}

The graph $\mathcal{G} = (\mathcal{V}, \mathcal{E})$ is constructed by representing each nucleus as a node. The node features are the learned latent representation of the nuclear cross sections obtained from the encoding module (either VAE and INR), i.e., vectors of dimension $k=32$. The edges $\mathcal{E}$ define the structural relationship between nuclei based on their positions in the $(Z,N)$ chart. Here, we employ 2-skip edge connectivity, where each node skips its 1-hop and 2-hop neighbors and connects to nodes located three steps away along the horizontal and vertical directions. This design choice is motivated by our goal to impute a relatively large contiguous missing region (e.g., a $9\times9$ test window). Using only local connectivity would limit the model's ability to propagate information into the interior of the masked test region. The designed skip connectivity enables longer-range message passing and allows the model to capture broader trends in the nuclear chart. We identified a performance trade-off between local dense connectivity and skip connectivity, particularly with respect to the size of the test window. For example, denser edge connections tend to yield better performance for smaller test windows, whereas a noticeable performance drop is observed when applied to larger test windows (e.g., $9\times9$), where longer-range dependencies become more critical.  

To effectively model such neighbor-dependent interactions, our GNN architecture $\chi(\cdot)$ leverages the \textit{message passing neural network} (MPNN) framework \cite{gilmer2017neural}, where the hidden state of each node is iteratively updated by aggregating information from its neighbors. Specifically, for the regular grid graph considered in this 
paper, the forward pass through the GNN has two phases---a message passing 
phase and a readout phase. In the message passing phase for a particular node, 
the message function $M_t$ aggregates the information from its neighbors and 
updates the node according to some update function $U_t$. At any time $t$, the 
hidden state $h_i^t$ at node $v_{i}$ is updated as
\begin{align}
m_i^{t+1} & =\sum_{j \in N(v_{i})} M_t\left(h_i^t, h_j^t, e_{i j}\right) \, ,\\
h_i^{t+1} & =U_t\left(h_i^t, m_i^{t+1}\right) \, ,   
\end{align}
where $N(v_{i})$ denotes the neighbors of node $v_{i}$ in the graph 
$\mathcal{G}$, and $e_{ij}$ represents an optional edge attributes. 


\subsubsection{GNN Training}
\label{subsubsec:training}

In our GNN-based imputation framework, training is performed by randomly masking a subset of nodes at each iteration and replacing their features with random placeholders. The GNN is then tasked with predicting the original values in the masked regions using the features of neighboring nodes and the graph structure. Importantly, the loss is computed only over the masked nodes. The test nodes (held-out $9\times9$ region centered at $(Z_0,N_0) = (42,60)$) are also replaced with fixed random placeholders, using a consistent random seed, and are excluded from training loss computations. 

We adopt two different training strategies depending on the underlying representation model, VAE or INR, used to generate node features. For VAE-based embeddings, we found that training the GNN to minimize reconstruction loss (Mean Square Error) in the original cross section space, i.e., between decoded predictions and ground truth, works well, particularly when the decoder is fine-tuned jointly with the GNN. This end-to-end training allows the decoder to adapt to the GNN's latent predictions, resulting in more accurate predictions. By contrast, for INR-based embeddings, we observe that freezing the INR decoder and minimizing the loss directly in the latent space leads to better performance. This might be because the INR decoder with hypernetwork that generates the weights of an implicit neural function makes it difficult to fine-tune effectively. Formally, the two training objectives read
\begin{align}
    \mathcal{L}_{\text{GNN/VAE}} & = \frac{1}{|\mathcal{M}|} \sum_{v_i \in \mathcal{M}} \left\| \phi\left( \chi(\mathcal{G})_i \right) - \mathbf{y}_i \right\|^2 \, ,\label{eq:loss_gnn_vae}\\ 
    \mathcal{L}_{\text{GNN/INR}} & = \frac{1}{|\mathcal{M}|} \sum_{v_i \in \mathcal{M}} \left\| \chi(\mathcal{G})_i - \mathbf{z}_{i} \right\|^2 \, .\label{eq:loss_gnn_inr}
\end{align}
Here, $\mathcal{M}$ denotes the set of masked training nodes. $\chi(\mathcal{G})_i$ denotes the GNN-predicted latent vector for node $v_i$, and $\phi$ represents the decoder that maps the latent vectors to the original cross section domain. $\mathbf{y}_i$ is the ground-truth cross section, while $\mathbf{z}_{i}$ refers to the corresponding ground-truth latent representation. The overall imputation pipeline is summarized in Algorithm \ref{alg:gnn_imputation}. Furthermore, we refer the reader to Appendix \ref{app:training_ablation} for a detailed comparisons for training strategies under different loss spaces and decoder fine-tuning options. 

\begin{algorithm}[H]
\caption{Graph-Based Imputation on Embeddings}
\label{alg:gnn_imputation}
\begin{algorithmic}[1]
\Require Node feature extractor \(\mathrm{E}\) (VAE or INR encoder), decoder \(\phi\) (VAE or INR), GNN \(\chi\), training set \(\mathcal{T}\), test region \(\mathcal{S}\) (e.g., \(9 \times 9\) window), mask probability \(p = 10\)\%, number of epochs \(T\)
\Ensure Predicted node features for \(\mathcal{S}\)

\State Encode raw cross section data: \(\mathbf{z}_i \gets \mathrm{E}(\mathbf{y}_i)\) for all \(v_i \in \mathcal{T} \cup \mathcal{S}\)
\State Build graph \(\mathcal{G} = (\mathcal{V}, \mathcal{E})\) with chosen edge connectivity (e.g., 2-skip edge connectivity)

\State Replace node features of all \(v_j \in \mathcal{S}\) with fixed random noise: \(\mathbf{z}_j \gets \eta_j\) with fixed seed

\For{\(t = 1\) to \(T\)}
    \State Sample a random subset \(\mathcal{M} \subset \mathcal{T}\) with \(|\mathcal{M}| = p \cdot |\mathcal{T}|\)
    \State Replace node features of \(\mathcal{M}\) with random noise \(\eta_i\)

    \State \(\hat{\mathbf{z}}_i \gets \chi(\mathcal{G})_i\) for all \(v_i \in \mathcal{V}\)

    \If{using VAE}
        \State \(\hat{\mathbf{y}}_i \gets \phi(\hat{\mathbf{z}}_i)\) \Comment{Decode latent}
        \State \(\mathcal{L} \gets \frac{1}{|\mathcal{M}|} \sum_{v_i \in \mathcal{M}} \left\| \hat{\mathbf{y}}_i - \mathbf{y}_i \right\|^2\) \Comment{Eq. (\ref{eq:loss_gnn_vae})}
    \Else
        \State \(\mathcal{L} \gets \frac{1}{|\mathcal{M}|} \sum_{v_i \in \mathcal{M}} \left\| \hat{\mathbf{z}}_i - \mathbf{z}_i \right\|^2\)  \Comment{Eq. (\ref{eq:loss_gnn_inr})}
    \EndIf

    \State Backpropagate and update parameters of \(\chi\) (and \(\phi\) if decoder is fine-tuned)
\EndFor

\State Use trained \(\chi\) (and \(\phi\)) to predict missing features for all \(v_j \in \mathcal{S}\)
\State Return \(\hat{\mathbf{y}}_j \gets \phi(\chi(\mathcal{G})_j)\)
\end{algorithmic}
\end{algorithm}


\subsubsection{GNN Architectures}
\label{subsubsec:prediction}

We evaluate three types of message-passing neural network architectures, ARMAConv \cite{bianchi2021graph}, GatedConv \cite{li2015gated}, and GraphSAGE \cite{hamilton2017inductive}. The ARMA convolutional  layer simulates graph convolution via approximated rational filters using a stack of recurrent units. It supports deeper feature propagation and has shown stability in capturing long-range dependencies. With shared weights across layers, it balances model expressiveness and parameter efficiency. The GatedConv integrates recurrent updates with gated mechanisms, allowing information to flow adaptively over multiple time steps. The gates help control feature propagation, which can be useful for graphs where information needs to be selectively passed or retained. The GraphSAGE generates node embeddings by aggregating features from a node's neighborhood. It supports scalable learning and provides robust performance across a variety of graph learning tasks. In our setup, it achieves the highest imputation performance with VAE embeddings while maintaining parameter efficiency. All these architectures considered use the MPNN backbone with slight variations either in terms of the type of filter used or the node update method, and each architecture is trained under the same data and masking setup.


\section{Results}
\label{sec:results}

We recall that our proposed framework consists of two main stages. First, we learn low-dimensional latent representations of nuclear cross sections using dimensionality reduction techniques. Second, we train GNN to impute the missing node features (i.e., latent vectors) in the nuclear chart. Since the effectiveness of GNNs in this task depends on the quality of embedding space, we begin by analyzing the two types of encoders used in this study, the VAE model of Fig.~\ref{fig:vae_inr}(a) and the INR model of Fig.~\ref{fig:vae_inr}(b). We then evaluate the performance of several GNN architectures for the downstream imputation task, where the goal is to predict unknown cross sections in the test region using information learned from known regions in the training set. 


\subsection{Encoding Performance}
\label{subsec:enc_perf}


To train models for latent representations, we randomly split the dataset containing $2171$ samples 
into a training set ($\approx$90\%) and a testing set ($\approx$10\%). To 
ensure reproducibility and fair comparisons, we employed a fixed random seed, 
guaranteeing both models were trained on identical datasets. Prior to model 
training, we normalized the input data by dividing each cross section by its 
corresponding maximum value across the entire dataset. This ensures that each 
cross section falls within the range of $[0,1]$. Then, the VAE model was 
trained on a multi-channel cross section dataset for $5,000$ epochs with a 
batch size of $32$. We utilized the Adam optimizer (initial learning rate 
$1 \times 10^{-3}$) for optimization. The INR model, trained on the same 
dataset, also used the Adam optimizer (learning rate $5 \times 10^{-5}$) but 
for $600$ epochs. Notably, the INR model demonstrated significantly faster 
convergence compared to the VAE training while maintaining high quality of 
latent representations.

\begin{figure}[!htb]
\centering
\includegraphics[width=\linewidth]{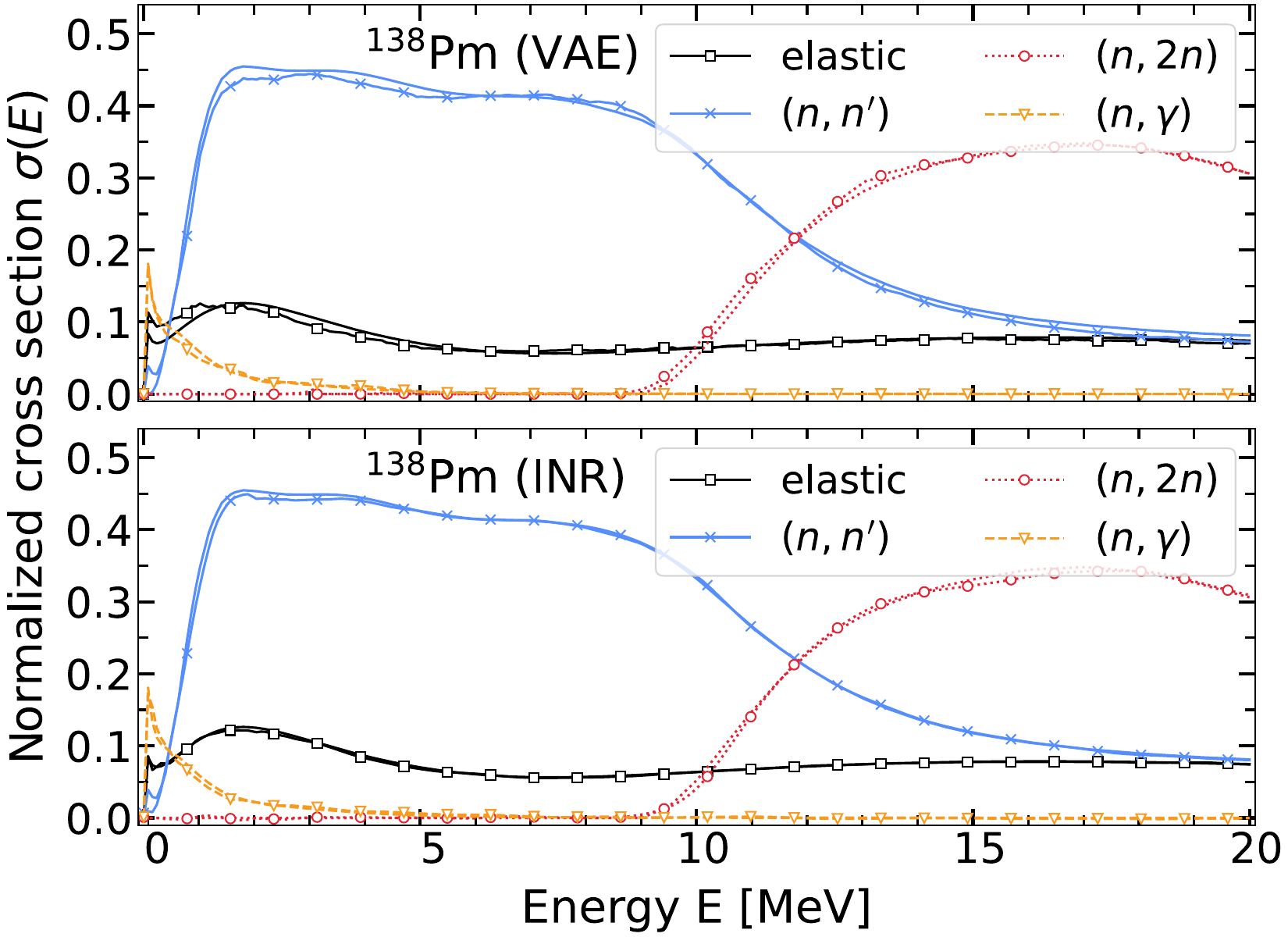}
\caption{Comparison of the encoding performance between VAE (top) and INR 
         (bottom) for all four cross sections in the $^{138}$Pm nucleus. 
         Ground-truth results are shown in plain lines and full symbols while 
         reconstructions are in dashed lines and open symbols. Performance is 
         typical of all nuclei in the TENDL dataset.}
\label{fig:recon_vae_inr}
\end{figure}

Figure \ref{fig:recon_vae_inr} illustrates the performance of both the 
VAE (top) and INR (bottom) for a nucleus within the TENDL dataset. Each color 
represents one of the four cross section channels considered in this work. The 
curves with symbols (square, circle, x, inverted triangle) represent the 
reconstructed inputs generated by either the VAE or the INR, whereas plain 
curves depict the ground truth inputs. As seen in this figure, both models 
effectively reconstruct given test samples, closely aligning with the ground 
truth. These results demonstrate that both models are able to encode the 
multi-channel cross section input into a compact, 32-dimensional latent vector 
with minimal loss in reconstruction accuracy. 

\begin{figure*}
\centering
\includegraphics[width=0.98\linewidth]{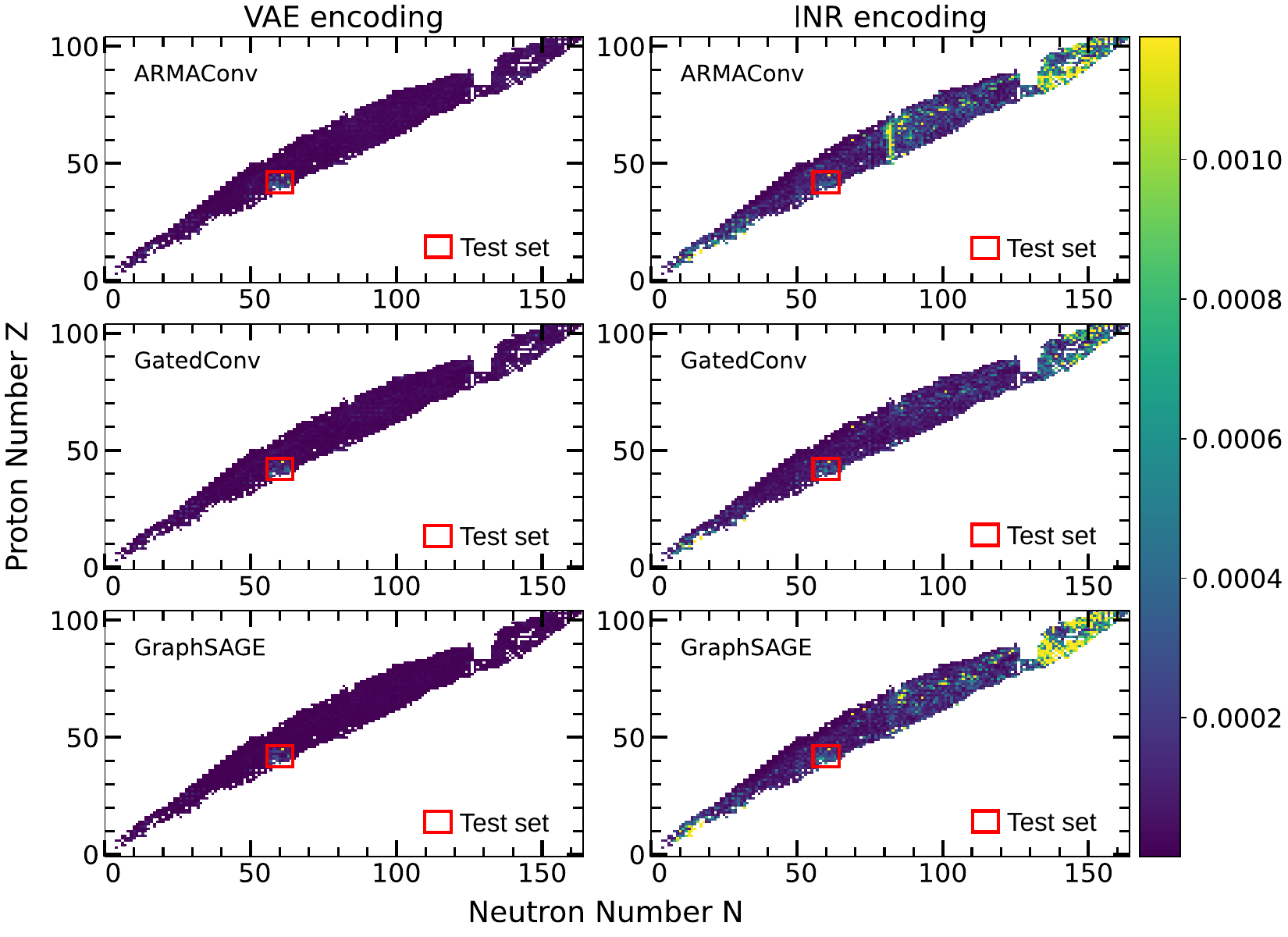}
\caption{Mean prediction errors calculated using Eq.~\eqref{eq:test_err} of the 
         GNN for each nucleus in the TENDL dataset. For better visualization, 
         all plots are normalized between $\epsilon_{\rm min}$ and 
         $0.03\times\epsilon_{\rm max}$, where $\epsilon_{\rm min}$ 
         ($\epsilon_{\rm max}$) is the minimum (maximum) error over all plots. 
         The three panels on the left side are based on VAE encoding while the 
         three on the right on INR encoding. ARMAConv, GatedConv and GraphSAGE 
         refer to the 
         three types of GNNs considered in this work. The red square 
         corresponds to the windowed test data: the GNN was not trained on it.
         }
\label{fig:windowed_heatmap_multi_channel}
\end{figure*}

To quantify reconstruction performance, we use the Peak Signal-to-Noise Ratio 
(PSNR). The PSNR is defined as follows
\begin{equation}\label{eq:psnr}
\text{PSNR} = 20 \log_{10}\left( \frac{\text{MAX}}{\sqrt{\text{MSE}}} \right) \, ,
\end{equation} 
where $\text{MSE} = \frac{1}{n}\sum^{n}_{i=1} 
\|\mathbf{y}_{i}-\hat{\mathbf{x}}_{i}\|_{2}^{2}$, $\hat{\mathbf{x}}_{i}$ 
represents the reconstructed input generated by either the VAE or the INR, 
while $\mathbf{y}_{i}$ denotes the original cross section inputs. Here, $n$ 
is the total number of testing samples. $\text{MAX}$ represents the maximum 
possible value of the signal, which is equal to $1$. We evaluate each model 
using this metric on the entire test set. The VAE achieves a score of $46.68$, 
while the INR achieves a higher score of $51.39$. Overall, the performance of 
the INR is significantly better as this model is better able to precisely 
capture high-frequency information such as thresholds, for example the fact 
that the cross section for the $(n,2n)$ cross section is identically 0 unless 
the energy of the incident neutron exceeds the neutron separation energy. 


\subsection{GNN performance}
\label{subsec:gnn_perf}

To evaluate the GNN performance, we divide the entire TENDL dataset into a training and testing set based on a windowing  function. The windowed test set is created using a square window centered around a specific nucleus $(Z_0,N_0)$ in the chart of isotopes and containing all nuclei with $Z_0 -\Delta Z \leq Z_0 \leq Z_0+\Delta Z$ and $N_0 -\Delta N \leq N_0 \leq N_0+\Delta N$. All results presented in this section were obtained using a test window centered at $(Z_0, N_0) = (42, 60)$ with $\Delta Z = \Delta N = 4$, resulting in a $9 \times 9$ test region on the nuclear chart. This window contains a total of 81 test nuclei, 67 of which are included in the original TENDL dataset. Since that original dataset contains 2171 points on  the 2D plane, this means that about 3.0\% of the total data is allocated to the  test set. In Appendix \ref{app:GNN}, we also present the GNN performance across varying window sizes and different center locations.

\begin{figure*}[!htb]
\centering
\includegraphics[width=\linewidth]{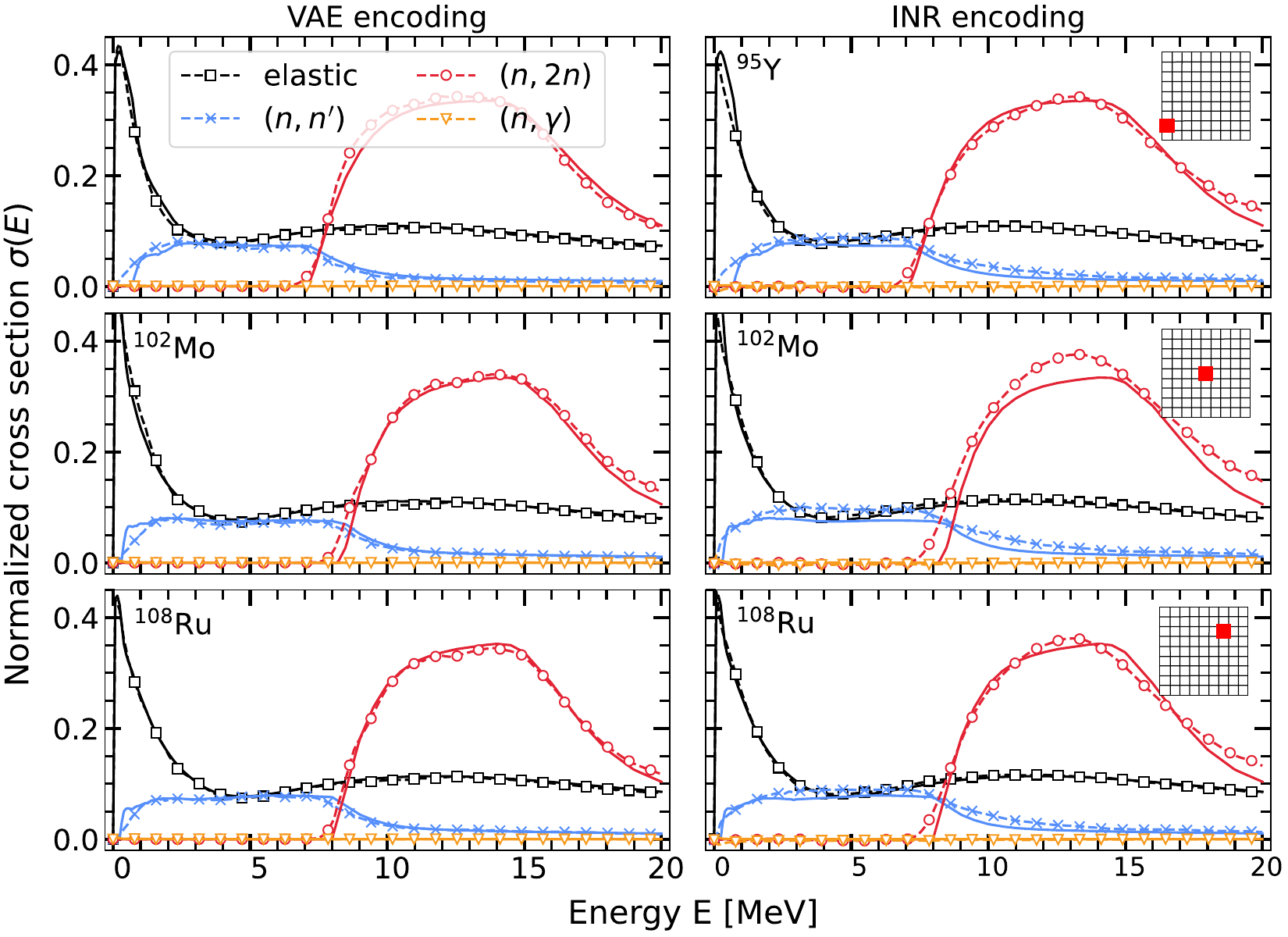}
\caption{Performance of the VAE (left) and INR (right) model for the GraphSAGE
         GNN architecture in predicting nuclear cross sections 
         within the test set for the central nucleus, $^{102}$Mo and the ones 
         near the lower left and upper right of the window, $^{95}$Y and 
         $^{108}$Ru respectively. Plain curves without symbols correspond to 
         the ground truth while dashed curves with open symbols correspond to 
         GNN predictions. All cross sections are normalized independently in 
         each channel across the entire dataset. The small grid in the upper 
         right corner of the INR figures shows the location of each nucleus 
         within the test set.}
\label{fig:multi_channel_test_pred}
\end{figure*}

To train the GNNs, we use the AdamW optimizer \cite{loshchilov2017decoupled} with a base learning rate of $lr=0.001$. To promote faster convergence and stabilize training dynamics, we adopt a OneCycle learning rate scheduler \cite{smith2018disciplined}, which linearly increases rate up to $10\times lr$ before annealing it back down. Specifically, we set the maximum learning rate to $0.01$, with a total of $100,000$ training epochs. This setup was applied consistently across both VAE-based and INR-based experiments. For each epoch, the entire training set is passed through the GNN and we do not create batches  of the training graph since breaking the edge relationships between the nodes would yield inaccurate results. The prediction error of the decoded cross sections is computed as: 
\begin{equation}
\label{eq:test_err}
    \epsilon = \frac{1}{n} \sum_{i=1}^{n} || \mathbf{y}_{i} - \mathbf{\hat{x}}_{i}||_{2}^2 \, ,
\end{equation}
where $\mathbf{\hat{x}}_{i}$ and $\mathbf{y}_{i}$ are the predicted node features and the ground truth in the original space, respectively, and $n$ is the actual number of nuclei in the set (either training or test). The quantity $(\mathbf{y}_{i} - \mathbf{\hat{x}}_i)$ computes the point-wise difference, which is then summed for each position $i$ in the test set. We calculate the prediction error in the 256-dimension space using Eq.~\eqref{eq:test_err}, and present the corresponding error distribution as a heat map in Fig.~\ref{fig:windowed_heatmap_multi_channel}.

In Fig.~\ref{fig:windowed_heatmap_multi_channel}, we observe notable differences in the error distribution between VAE and INR embeddings across the nuclear chart. In the VAE case (left column), prediction errors are mostly concentrated in the test region, while the rest of the chart (training region) shows low errors. Conversely, INR-based models (right column) display more widespread errors, including in the training region, especially in the upper part of the chart. This aligns with INR's relatively compact latent space as well as lower diversity across samples, which leads to a smaller training-testing gap. We further examine and visualize the characteristics of these latent spaces in Section \ref{subsec:latent_vis}. To provide a more quantitative comparison, Table \ref{tab:DA-noDA-comparison} reports the average test PSNR of Eq.~\eqref{eq:psnr} over all test samples for each GNN architecture using both VAE and INR embeddings.

\begin{table}[!htb]
\caption{Test PSNR values of the different GNN architectures with VAE and INR embeddings.}
\label{tab:DA-noDA-comparison}
\begin{ruledtabular}
\begin{tabular}{lccc}
    GNN Arch.        & ARMAConv  & GatedConv & GraphSAGE \\ \hline
INR embedding   & 35.31 & \textbf{35.48} & 35.14\\
VAE embedding & 35.87 & 36.40 & $\mathbf{36.65}$ \\
\end{tabular}
\end{ruledtabular}
\end{table}

We also visualize the prediction performance of the GraphSAGE network using both VAE-encoding (left column) and INR-encoding (right column) in Fig.~\ref{fig:multi_channel_test_pred}. The figure shows the normalized cross sections for three nuclei located within in the test window: $^{95}$Y near the lower-left corner, $^{102}$Mo at the center (hence the 
furthest away from the training set), and $^{108}$Ru near the upper-right corner of the window. With the VAE-encoding, the predicted curves closely match the ground truth, with only a slight deviation in the $(n,2n)$ channel. The predictions from the INR-encoding exhibit similar trends and capture the overall structure of the cross sections, despite more noticeable differences in $(n,2n)$ channel, especially at the center region.  
Numerical experiments show that the model's epistemic error stemming from the 
training of both VAE/INR and GNN is small and have no effect on these conclusions.
Note that the $(n,\gamma)$ channels is not identically zero but the normalized cross section is several order smaller than 
the other ones for these particular nuclei.


\section{Discussion} 
\label{sec:discussion}

The results presented in Section \ref{sec:results} pose a few interesting 
questions, both from a data science and from a nuclear physics perspective, 
which we discuss in this section. In particular, we focus on a brief analysis 
of the structure of the latent space to analyze whether specific patterns can be 
matched to nuclear properties. 


\subsection{Representation learning and structure of the latent space} 
\label{subsec:latent_vis}

\begin{figure}[!htb]
\centering
\includegraphics[width = 0.95\linewidth]{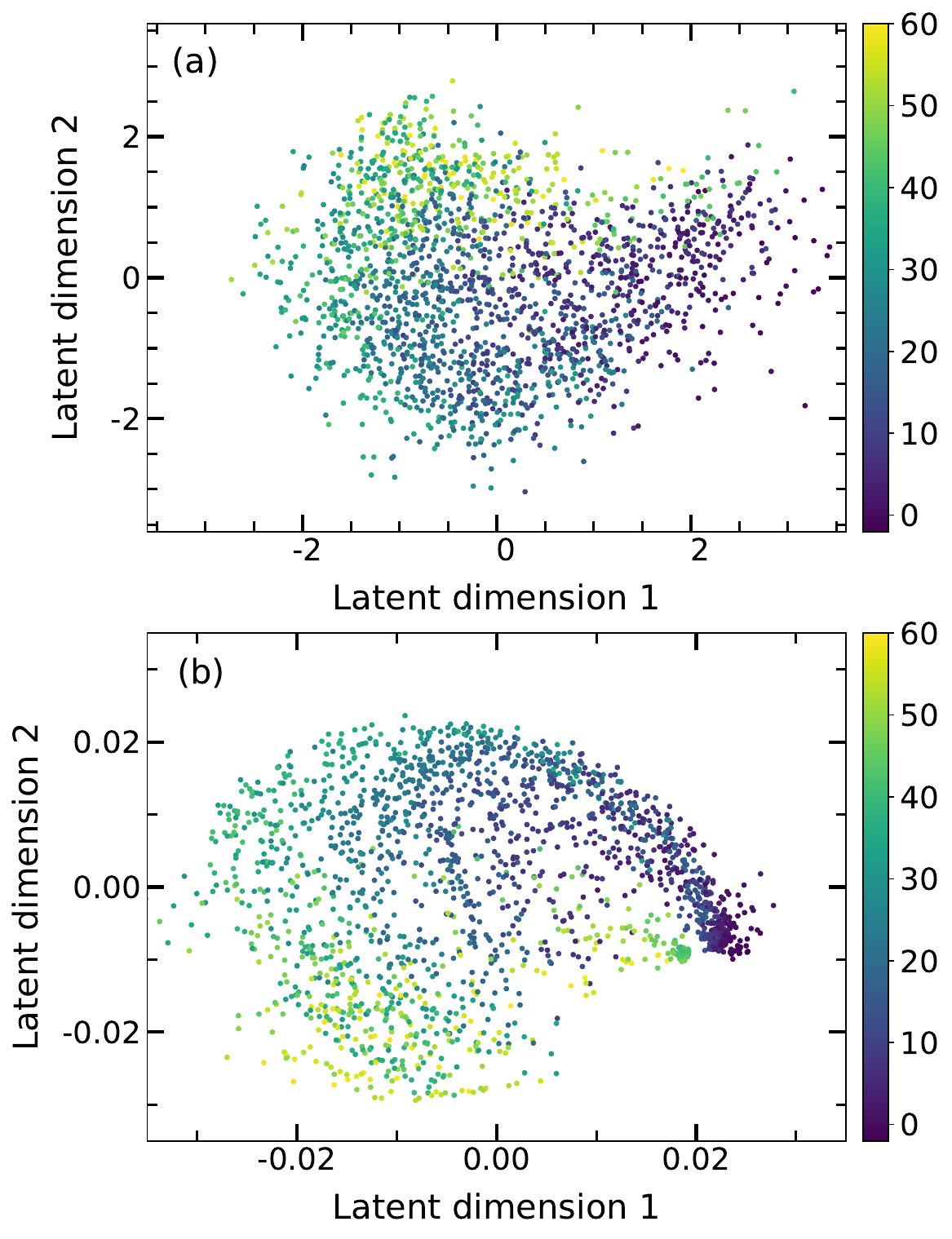}
\caption{PCA visualization of the VAE and INR hypernet latent spaces. 
         Panel (a): projections of each $k$-dimensional VAE latent vector onto 
         a two-dimensional plane spanned by the two largest components of the 
         PCA. Panel (b): same figure for the INR latent vector.}
\label{fig:latent_PCA}
\end{figure}

When training GNNs with different latent representations, the choice of the loss function \eqref{eq:loss_gnn_vae} or \eqref{eq:loss_gnn_inr} significantly impacts performance. Our experiments reveal a clear contrast: GNNs trained on VAE-based embeddings achieve better performance when optimizing in the original cross section space, whereas GNNs trained on INR-based embeddings perform better when optimized directly in the latent space. This discrepancy can be attributed to two key factors: 
\begin{description}
\item[Decoder sensitivity in INR models] 
The INR decoder is parameterized by a hypernetwork which maps latent vectors not directly to outputs but to the weights of an implicit neural function. This architectural design introduces an additional sensitivity when it is used as downstream prediction task. Small perturbations in the predicted weights of the neural representation can drastically affect the predicted cross sections. This makes finetuning the decoder in an end-to-end fashion less effective compared to the one with VAE-based embedding. 
\item[Embedding space expressivity] 
As shown in Fig.~\ref{fig:latent_PCA}, VAE latent spaces tend to have higher variance and broader spread across samples, which provide the downstream GNN with richer signals to learn from, whereas, INR latent spaces are more compact and concentrated in narrow regions of the latent space. This could limit the discriminative power of the GNN. Consequently, optimizing directly in the latent space for INR-based embedding circumvents the difficulty of optimization through the sensitive decoder. 
\end{description}


In Fig.~\ref{fig:latent_PCA}, we show a direct visualization of the latent 
spaces for the training dataset of both the VAE and INR models. In both panels, 
we perform a principal component analysis (PCA) of the $k$-dimensional latent 
vectors ($k=32$) and project each of them onto the two-dimensional plane 
spanned by the two largest PCA components. Panel (a) corresponds to the VAE 
latent space while panel (b) corresponds to the INR latent space. We recall 
that each point in such a figure corresponds to a given nucleus; its color is 
the value of the difference $N-Z$ of the neutron number $N$ and proton number 
$Z$ in the corresponding nucleus. The only noticeable difference is that the 
samples' variability scale for the latent space in the INR model is 
significantly smaller than for its VAE counterpart: the INR hypernet latent 
space is much more ``compact'' than the VAE one.

\begin{figure}[!htb]
\centering
\includegraphics[width = 0.91\linewidth]{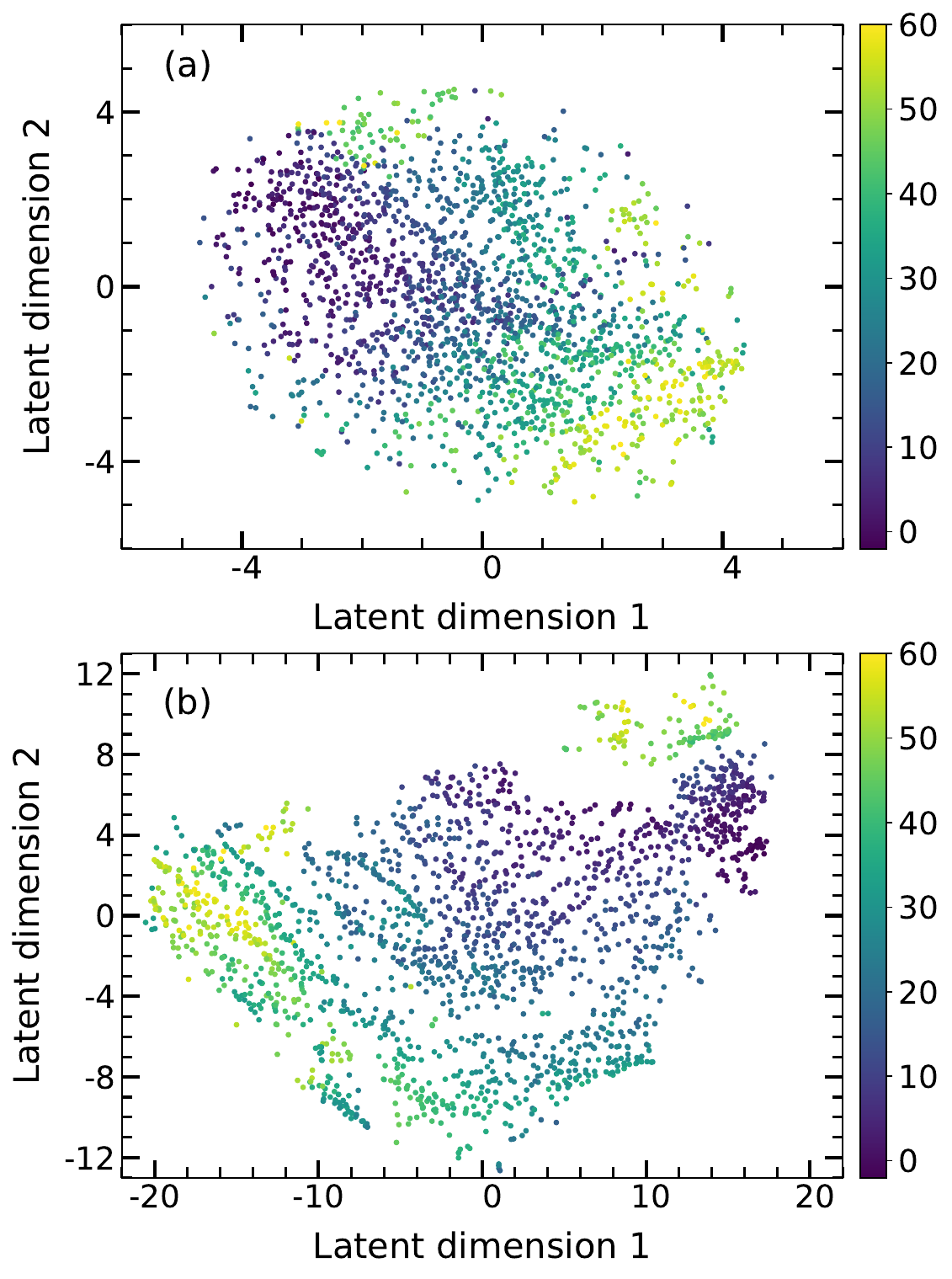}
\caption{t-SNE visualization of the VAE and INR hypernet latent spaces. Panel 
         (a): two-dimensional t-SNE embedding of $k$-dimensional VAE latent 
         vectors obtained for a perplexity value of 200. Panel (b): same figure 
         for the INR latent vector.}
\label{fig:latent_TSNE}
\end{figure}

To better understand the structural differences between the two encoding 
techniques, we compute the t-distributed stochastic neighbor (t-SNE) embedding 
\cite{maaten2008visualizing}. Let us quickly recall the basics of the t-SNE 
method. Assume we have a set $\mathcal{P}$ of points $P$ in a $n_p$-dimensional
space, and characterize the ``similarity'' of two points $P$ and $P'$ in 
$\mathcal{P}$ by some joint probability distribution function $p$. Let us now 
assume that $\mathcal{P}$ can be mapped onto a set $\mathcal{Q}$ of points $Q$ 
in a $n_q$-dimensional space ($n_q \ll n_p$) that is characterized by another 
probability distribution $q$. The t-SNE technique minimizes the 
Kullback-Leibler divergence $\mathcal{L}_{PQ}$ between the joint probabilities 
$p$ in the high-dimensional space and the joint probabilities $q$ in the 
low-dimensional space. In simple terms: two points that are ``neighbors'' in 
the high-dimensional space should also be neighbors in the low-dimensional one.

Figure \ref{fig:latent_TSNE} shows the two-dimensional representation of this 
t-SNE embedding for both the VAE and INR encodings as obtained with the 
{\tt scikit-learn} implementation. Like Fig.~\ref{fig:latent_PCA}, the color of 
a point is the difference $N-Z$ of the neutron number $N$ and proton number $Z$ 
in the corresponding nucleus. In practice, t-SNE algorithms require setting a 
few hyperparameters, most notably one called ``perplexity'' $\perplex$ which 
can be interpreted as ``{\em a smooth measure of the effective number of 
neighbors}'' in the high-dimensional space \cite{maaten2008visualizing}. 
Although traditional applications of t-SNE suggest values of the perplexity 
between 5 and 50, this is highly dependent on the dataset and its 
dimensionality. In our case, the dataset has 2090 points---training nuclei 
represent 97\% of the whole TENDL dataset of 2171 nuclei---and is encoded in a 
latent space of dimension $k=32$. We found that in order to converge the 
Kullback-Leibler loss $\mathcal{L}_{PQ}$, the perplexity parameter had to be 
significantly larger: Fig.~\ref{fig:latent_TSNE} was obtained for 
$\perplex = 200$. Note that the spatial arrangement of points and the 
variability scale in each latent dimension are very dependent on the value of 
$\perplex$. In particular, small values of $\perplex$ (corresponding to large 
values of $\mathcal{L}_{PQ}$) lead to excessive visible clustering while failing 
to converge the projection. 

\begin{figure}[!htb]
\centering
\includegraphics[width = 0.95\linewidth]{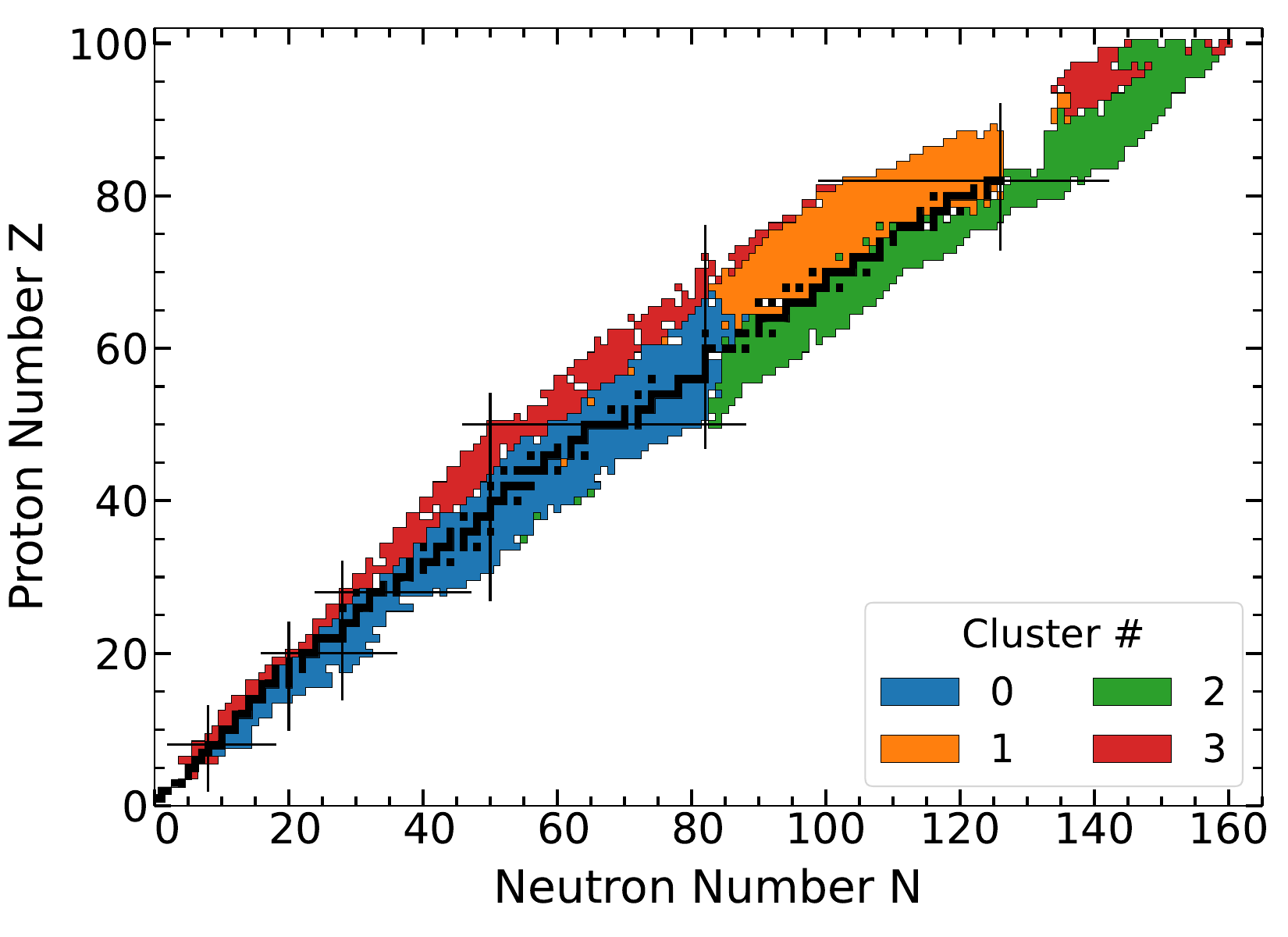}
\caption{Location of each nucleus in the training set with color code based on 
         which of 4 clusters the nucleus belong too. Black horizontal and 
         vertical lines represent magic numbers. Black squares are stable nuclei.}
\label{fig:latent_location}
\end{figure}

To better analyze clustering structures, we used the {\tt scikit-learn} 
implementation of the spectral clustering algorithm to identify potential 
clusters of points directly in the VAE or INR dataset. Out of all available 
clustering techniques, spectral clustering and the DBScan algorithms are the 
naive best choices to identify clusters based on existing documentation and 
literature. However, calibrating the maximum distance between any two points to 
be considered in the neighborhood of each other in DBscan requires a deeper 
understanding of the latent space than we currently have. In contrast, spectral 
clustering does not involve any prior knowledge of the latent space. We first 
determined a rough estimate of the optimal number $\nclus$ of clusters based on 
the Silhouette method. In practice, we found $\nclus = 4$. Figure 
\ref{fig:latent_location} shows the position in the nuclear chart of each 
nucleus from the INR encoding training dataset, color-coded by the cluster the 
nucleus belongs to. The most striking feature of this plot is the fact that 
nuclei are organized in diagonal bands loosely following the valley of 
stability rather than along isotopic, isotonic or isobaric sequences. In 
addition, some of the clusters seem to be related to neutron magic numbers 
(but, interestingly, not proton magic numbers), especially $N=82$ and $N=126$. 
Both observations are quite robust. In Appendix \ref{app:TSNE}, we show that 
different clustering scenarios lead to similar qualitative results.


\subsection{Predictions of unknown cross sections} 
\label{subsec:physics}

Predicting nuclear cross sections across the entire chart in a consistent 
manner and with well-defined uncertainties is particularly challenging due to 
the lack of experimental data on radioactive isotopes and the limited 
predictive power of existing models. In Section \ref{sec:results} we have shown 
that both VAE and INR networks can successfully compress four-channel cross 
sections with full energy index into compact, low-dimensional representations. 
This approach exhibits excellent encoding performance. This non-trivial result 
suggests that the ``effective'' space of nuclear cross sections is 
low-dimensional indeed.   

The predictive performance of GNNs in our imputation task is influenced by several interconnected factors. One of the most critical elements is the test window size, which directly impacts the number of available training nodes in the transductive setting where the model is trained and evaluated on a fixed global graph. As the size of the test window increases, fewer nodes remains for training, thereby limiting the opportunity for the model to learn from surrounding neighborhoods. This degradation in performance with larger test windows is confirmed in Fig.~\ref{fig:test_psnr_window}. Furthermore, nuclei occupy a diagonal band in the $(Z,N)$ chart rather than a uniform 2D grid. This results in regions of sparsity and disconnection that challenge the GNN's ability to propagate information effectively. As a result, the location of the test window also significantly affects performance, for example, regions with denser nuclear neighborhood exploit more effective message passing, whereas sparse areas hinder it as verified in Fig.~\ref{fig:test_psnr_regions}. In terms of training strategy, our results show that GNN performance depends not only on the structure of the latent space learned through VAEs or INRs but also on how the GNN is optimized. Specifically, the GNN with VAE embeddings benefit from end-to-end training in the original cross section space, while INR embeddings perform better when optimization is restricted to the latent space as reported in Table \ref{app:loss_ablation}. Our analysis from mulitple perspectives, including test PSNR, error heatmaps, and the structure of the learned latent space, highlights the critical importance of the embedding in determining the effectiveness of downstream GNN predictions. These findings suggest that developing latent representations that are well aligned with the needs of graph-based prediction tasks is a promising direction for future research. 


The obvious limitation of this initial study is that the training of all 
networks is based on the TENDL dataset, which is itself a tabulation of 
predictions from a finite collection of nuclear models. However, this could be 
turned into an advantage: since the deep learning approach provides a single 
unifying representation of all cross sections in that set, cross sections that 
the GNN fails to reproduce and/or predict well may be used as signals for 
outliers in TENDL. Furthermore, our methodology can easily be extended by 
augmenting TENDL with additional datasets, either from evaluated libraries 
such as ENDF, JEFF or JENDL, or from pure experimental data such as EXFOR. 

Most importantly, the learned relationships between cross sections of different
types within a single nucleus, or between different nuclei can pave the way to 
much more ambitious studies. In the short term, short-distance extrapolation 
of cross sections of the type 
$\sigma_{\alpha}(N, Z) \rightarrow \sigma_{\alpha}(N+i, Z+j)$ for channel 
$\alpha$ should be relatively reliable, especially if $i$ and $j$ are small, 
i.e., $i, j \leq 4$. Such extrapolations would be especially useful in the 
context of nucleosynthesis simulations where precision matters less than 
accuracy. In the longer term, it should be possible to exploit these 
relationships to estimate covariances. Current libraries are severely limited 
in covariance data and in many cases, these data are unreliable. Furthermore, 
cross-material covariances, that is, how uncertainties in a cross section of 
one nucleus are related to another type of cross section in another nucleus, 
have never been estimated. By providing a unified framework to compute cross 
sections, our methodology offers a viable path forward to estimate all such 
covariances and properly propagate uncertainties from nuclear data into 
applications. In the next section, we show how the current trained networks 
can already give us valuable information how uncertainties propagate across the 
nuclear chart.

\section{Conclusion} 
\label{sec:conclusion}

In this work, we used variational autoencoders (VAEs) and implicit neural 
representations (INRs) to learn the latent space of nuclear cross sections, 
specifically the elastic, inelastic, $(n,2n)$ and $(n,\gamma)$ cross sections 
in the fast regime. Both types of architecture were trained on the synthetic 
data from the TENDL nuclear dataset. We then used graph neural networks (GNNs) 
on the latent representations to predict the missing cross sections across the nuclear chart by leveraging the topological structure. 
We showed that we could predict cross sections of a rather decent-sized test set with good accuracy using a physics-free methodology: 
the networks learn from the data, whether the data come from measurements, 
calculations or evaluations. The fact that this proof-of-concept study used 
the TENDL dataset was motivated exclusively by the need to have enough data 
to train the networks.

Our initial results suggest that modern deep learning methods are capable of 
accurately predicting, in a single, unifying framework, different types of 
nuclear cross sections. This opens the way to a more rigorous quantification of 
uncertainties, especially covariances and cross-material covariances. Since 
experimental data on nuclear cross sections are rather scarce (at the scale of 
the chart of isotopes), it should be possible to augment datasets with other 
sources of data coming, e.g., from large-scale simulations of ground-state 
properties, $\beta$-decay rates and $\gamma$-strength functions, or level 
densities. Microscopic theories such as, e.g., nuclear density functional 
theory, should have enough predictive power to predict accurately the trends of 
these various observables at the scale of the mass table, which could in turn 
help constrain deep learning models during the training phase.


\section*{Acknowledgements}

This work was performed under the auspices of the U.S. Department of Energy by 
Lawrence Livermore National Laboratory under contract DE-AC52-07NA27344.
Lawrence Livermore National Security, LLC. This work was supported by the 
Laboratory Directed Research and Development (LDRD) program under project 
tracking code 23-SI-004. Computing support came from the Lawrence Livermore 
National Laboratory (LLNL) Institutional Computing Grand Challenge program.


\section{Appendix}
\label{sec:appendix}

This Appendix contains supplementary analyses and ablation studies to 
support the findings presented in the main paper. We first begin by investigating the representation learning phase, followed by additional analyses of the GNN-based prediction task, including variations in the test region location, test window size, and different training strategies.


\subsection{Train-Test Performance Trends for VAE}
\label{app:VAE}

\begin{figure}[!htb]
\centering
\includegraphics[width=1.0\linewidth]{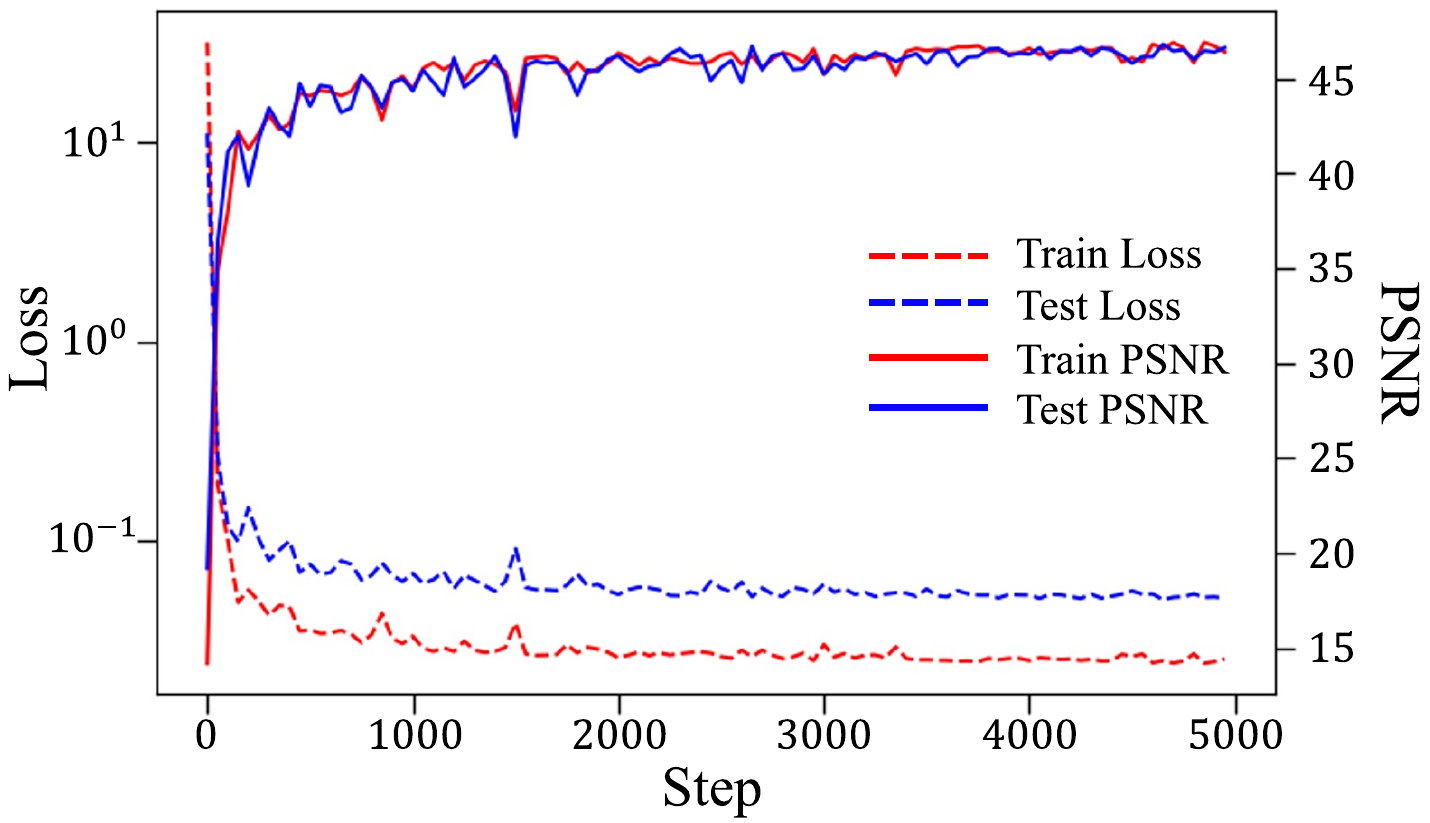}
\caption{Loss (dashed curves) and PSNR (plain curves) for the training set 
         (red) and testing set (blue) as a function of epoch during VAE 
         training. Note that the y-axis for the loss is on a logarithmic 
         scale.}
\label{fig:vae_log_trend}
\end{figure}

To ensure the model does not overfit during the representation learning phase, we analyze the evolution of the loss and PSNR for both the training and 
testing set during the VAE training. These metrics are presented as a function 
of epochs in Fig.~\ref{fig:vae_log_trend}. Both training and testing loss 
curves steadily decrease and converge to similar values by the end of 
training. Also, the PSNR values for both training and testing increase 
consistently, improving reconstruction quality over time. The consistent 
convergence of both metrics indicates minimal overfitting and robust 
generalization of the learned latent representations. 

\subsection{Generalization Analysis in GNN}
\label{app:GNN}

\begin{figure}[!htb]
\centering
\includegraphics[width=1.0\linewidth]{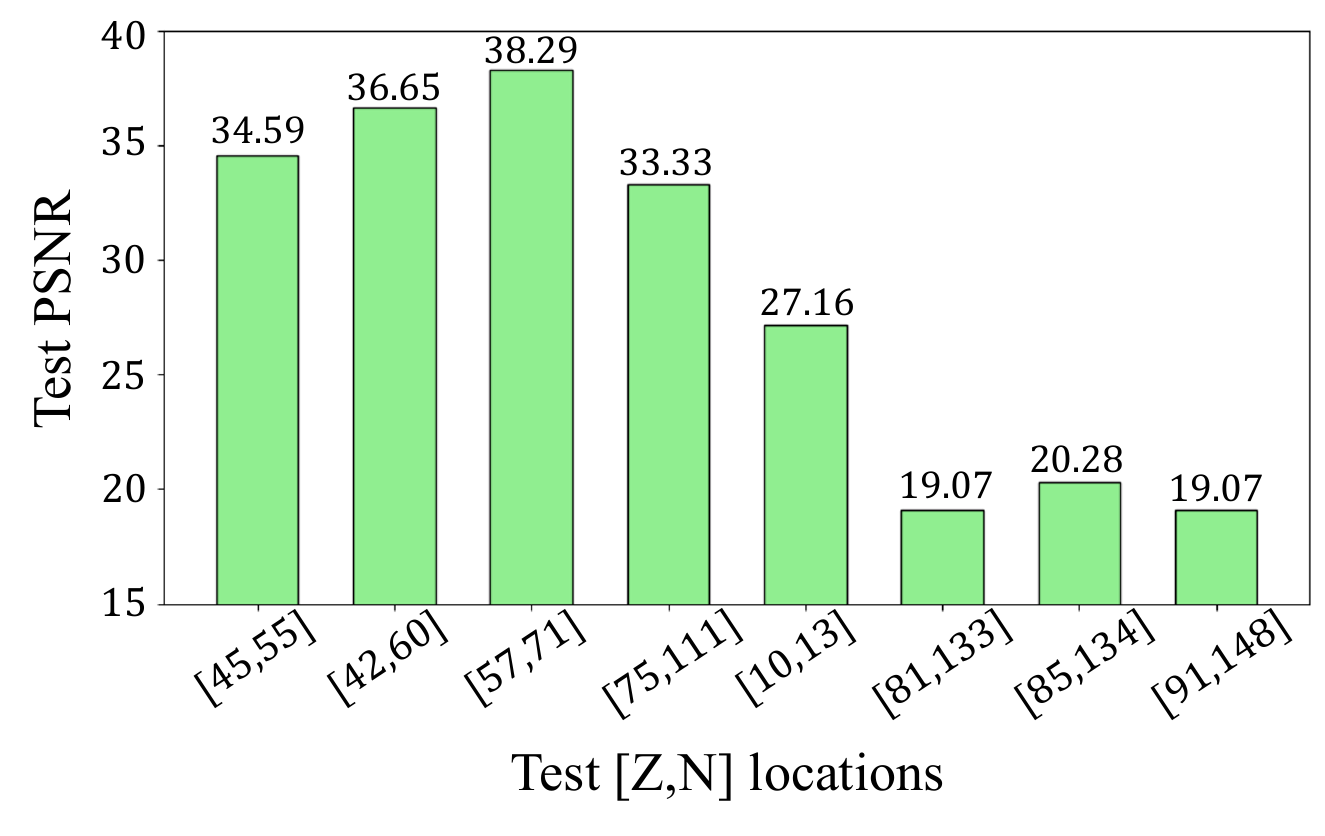}
\caption{Test PSNR scores for GNN imputation in test regions centered at different $(Z_0, N_0)$ across the nuclear chart. The performance trend reflects the influence of local structural support on message-passing effectiveness and prediction quality. All test regions have the same size characterized by $\Delta Z = \Delta N = 4$ (i.e., $9 \times 9$ test window); only the center nucleus changes.}
\label{fig:test_psnr_regions}
\end{figure}

\paragraph{Choice of test regions} 
To investigate the model performance in imputation across different regions of the nuclear nuclear chart, we evaluate the test PSNR scores at eight distinct locations $(Z_0,N_0)$ for the center of the test region. These locations vary in their local neighborhood density. For example, test regions centered on medium-mass nuclei such as $(Z_0,N_0)=(45,55)$, $(Z_0,N_0)=(42,60)$, and $(Z_0,N_0)=(57,71)$ are densely surrounded by known neighbors, while test regions centered on heavy nuclei such as $(Z_0,N_0)=(81,133)$, $(Z_0,N_0)=(85,134)$, or $(Z_0,N_0)=(91,148)$ are closer to the boundaries of the domain and are surrounded by fewer neighbors; see Fig.\ref{fig:latent_location} for example. As shown in Fig.~\ref{fig:test_psnr_regions}, the imputation performance is strongly correlated with local neighborhood density. Test regions with dense neighboring support yield higher PSNR scores, indicating effective message passage and generalization. In contrast, edge regions with limited topological context yield noticeably lower PSNR, reflecting the increased difficulty of inference. This observation highlights the importance of neighborhood context in graph-based models and the advantages of transductive message-passing when applied within densely populated nuclear regions.

\begin{figure}[!htb]
\centering
\includegraphics[width=0.9\linewidth]{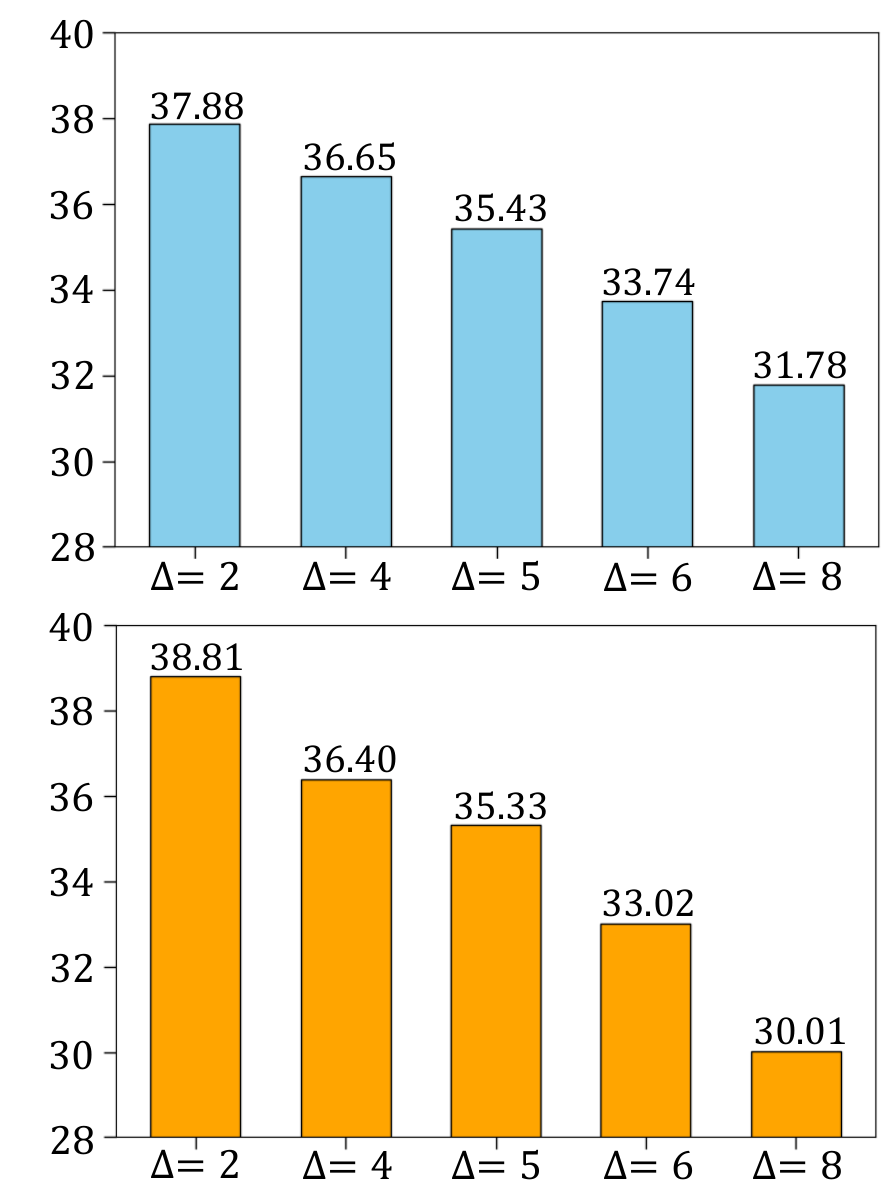}
\caption{Test PSNR values across varying window sizes for GraphSAGE (left) and GatedConv 
         (right) approaches, with the center of the test window fixed at 
         $(Z_0,N_0) = (42,60)$. The 2-skip edge connectivity is used for all cases.}
\label{fig:test_psnr_window}
\end{figure}

\paragraph{Choice of test window lengths} 
To further investigate how the imputation performance is influenced by the spatial extent of the test region, we analyze the GNN model's performance across various test window sizes denoted by $\Delta$. In this setup, a test window of size $\Delta \times \Delta$ is held out and masked during training, while all other surrounding data points are used for learning. All windows are centered on our default ($Z_0,N_0) = (40,62)$ nucleus. Since our graph is built based on spatial nuclear chart connectivity, increasing $\Delta$ reduces the number of available training nodes, making the imputation task increasingly challenging. As shown in Fig.~\ref{fig:test_psnr_window}, both GraphSAGE and GatedConv exhibit a clear degradation in test PSNR as the window length increases from $\Delta=2$ to $\Delta=8$. This trend aligns with the intuition that accurate imputation demands more effective long-range message passing for larger masked regions. This result highlights the importance of carefully balancing test window size and graph connectivity design. This is especially relevant in nuclear physics applications where the finite extent of the nuclear chart combined with the limited availability of actual data to train poses practical constraints.


\subsection{Loss Space and Decoder Fine-tuning Effects} \label{app:training_ablation}
In this section, building on the training discussion in Section \ref{subsubsec:training}, we present an ablation study comparing different optimization strategies for VAE-based embeddings and INR-based embeddings. Specifically, we analyze the effect of minimizing the loss in either the latent space (dimension: $k=32$) or the original space (dimension: $m=256$), with and without decoder fine-tuning. Let $\mathcal{L}_{\text{GNN/32}}$ denote the MSE loss computed between predicted latent vectors and corresponding ground-truth latent vectors, and let $\mathcal{L}_{\text{GNN/256}}$ denote the MSE loss computed between predicted and ground-truth cross sections in the original input space. The result in Table \ref{app:loss_ablation} highlights the distinct behaviors between the VAE and INR embeddings under different training regimes. When minimizing the loss in the original input space, the VAE-based embedding achieves the highest PSNR score when the decoder is fine-tuned jointly with the GNN. This suggests that the VAE's decoder is flexible enough to adapt to latent predictions during training, allowing for better end-to-end optimization. Interestingly, INR embeddings perform best when trained with loss computed directly in the latent space without involving the decoder in the optimization loop. This setup bypasses the complications associated with fine-tuning the INR hypernetwork. These observations justify the different training strategies outlined in Algorithm \ref{alg:gnn_imputation} for each type of embedding.

\begin{table}[!htb]
\caption{Comparison of test PSNR scores under different training strategies with GatedConv. $\mathcal{L}_{\text{GNN/256}}$: MSE in original cross section space; $\mathcal{L}_{\text{GNN/32}}$: MSE in latent space. Note that for $\mathcal{L}_{\text{GNN/32}}$ (Decoder ON), the decoder does not affect the optimization process as the loss is computed before decoding.}
\label{app:loss_ablation}
\begin{ruledtabular}
{\begin{tabular}{lcccc}
 & \multicolumn{2}{c}{Decoder OFF} & \multicolumn{2}{c}{Decoder ON} \\
\hline
Loss Type & INR & VAE & INR & VAE \\
\hline
$\mathcal{L}_{\text{GNN/256}}$ & 25.85 & 35.78 & 34.52 & \textbf{36.40} \\
$\mathcal{L}_{\text{GNN/32}}$  & \textbf{35.48} & 33.21 & -- & -- \\
\end{tabular}}
\end{ruledtabular}
\end{table}


\subsection{Location of latent clusters}
\label{app:TSNE}

\begin{figure}[!htb]
\centering
\includegraphics[width = 0.95\linewidth]{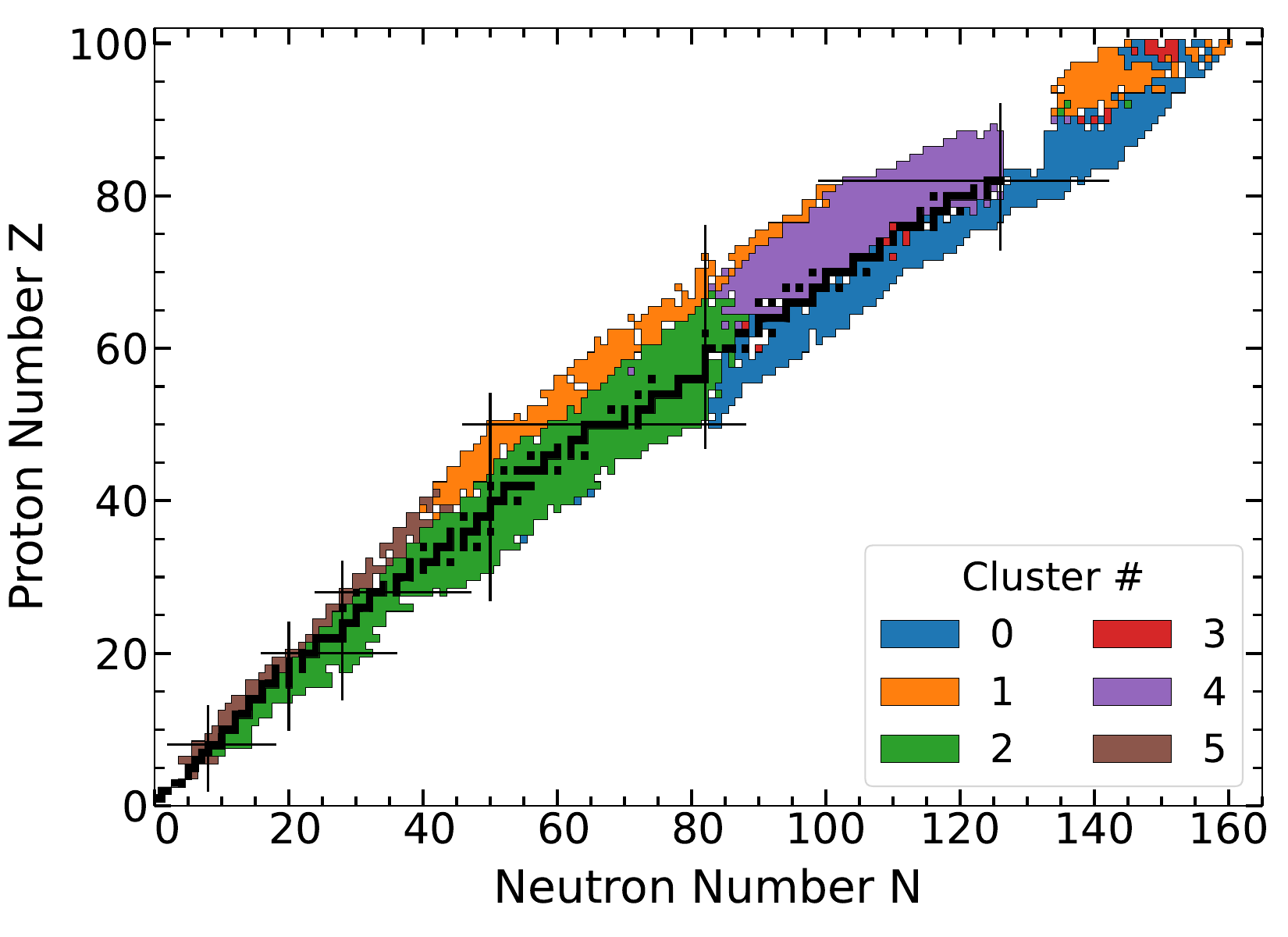}
\caption{Location of nuclei in the training set with color code based on which 
         cluster the nucleus belongs too. Black horizontal and vertical lines 
         represent magic numbers. The figure corresponds to the INR latent 
         vectors and: $\nclus=6$.}
\label{fig:latent_location_1}
\end{figure}

In this section, we show a few more examples of the location of clusters of 
nuclei identified directly from the VAE or INR latent vectors. Figures 
\ref{fig:latent_location_1} quantifies the impact of changing the numbers of 
clusters to $\nclus = 6$.  The main change is the splitting of the cluster of 
proton-rich nuclei into two subsets and the appearance of a small cluster of 
very heavy nuclei.

\begin{figure}[!htb]
\centering
\includegraphics[width = 0.95\linewidth]{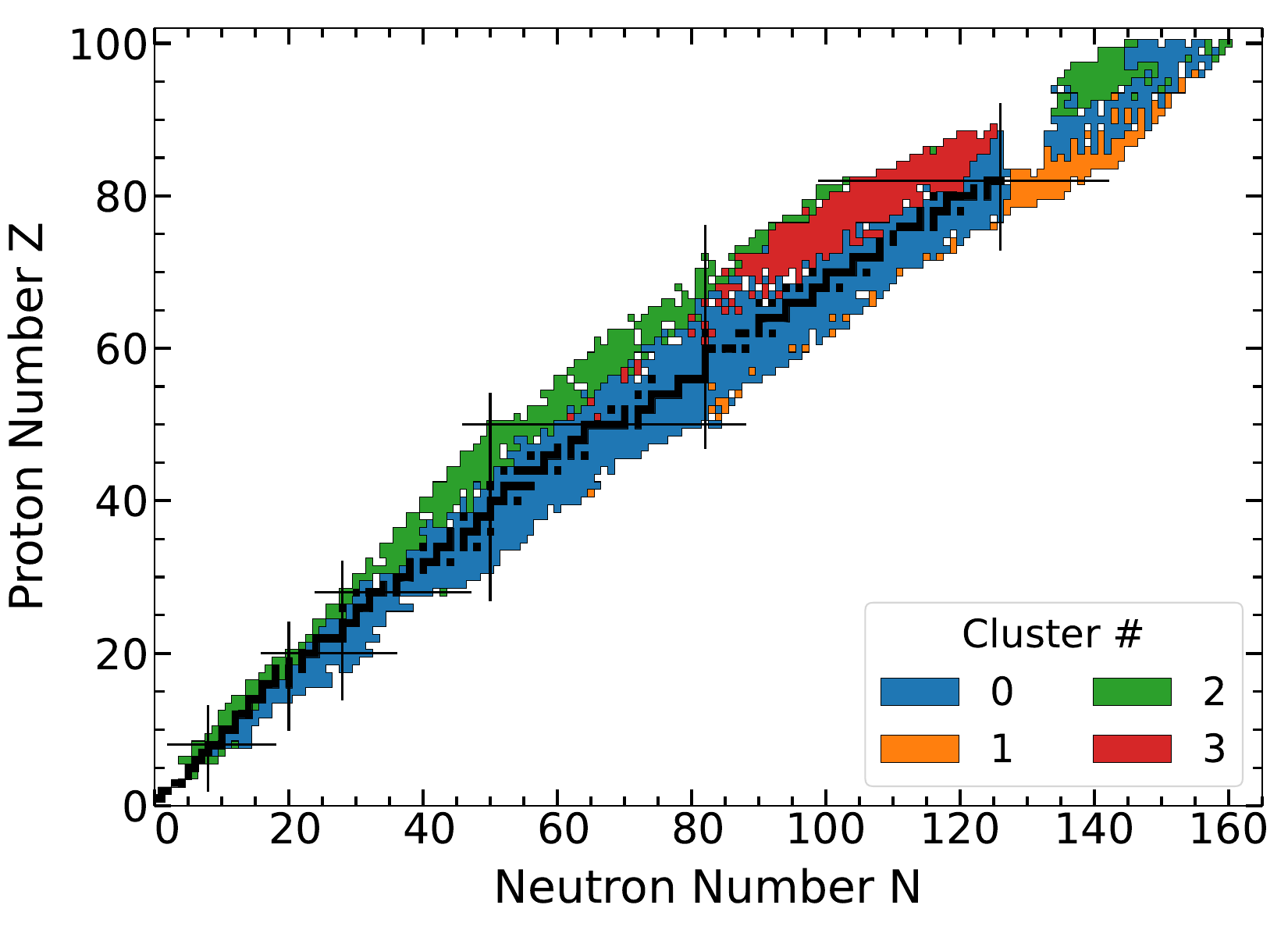}
\caption{Same as Fig.\ref{fig:latent_location_1}. The figure corresponds to the 
         VAE latent vectors and: $\nclus=4$.}
\label{fig:latent_location_2}
\end{figure}

Figure \ref{fig:latent_location_2} is the exact same figure as 
Fig.\ref{fig:latent_location} in the main text, only it was constructed from 
the spectral clustering of the VAE latent space (with $\nclus=4$). We notice 
that the overall ``diagonal'' grouping of nuclei remains, although clusters are 
a little less compact. In addition, the main cluster that surrounds the valley 
of stability is not split in two at $N=82$ like in 
Fig.~\ref{fig:latent_location} but instead at $N=126$. It is still interesting 
that neutron magic numbers seem to be identified as markers of clustering in 
both methods.

\begin{figure}[!htb]
\centering
\includegraphics[width = 0.95\linewidth]{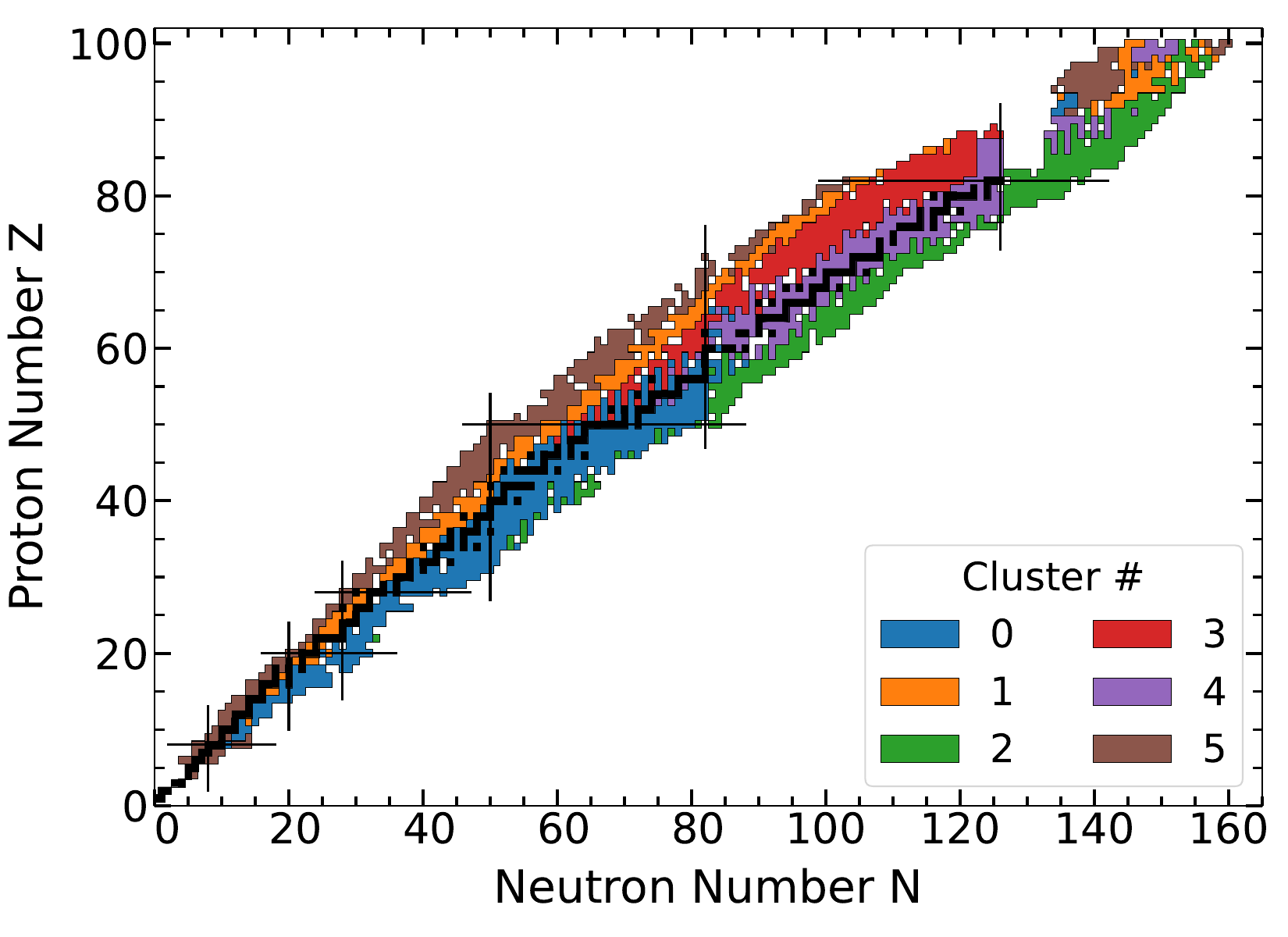}
\caption{Same as Fig.\ref{fig:latent_location_1}. The figure corresponds to the 
         INR latent vectors and is obtained by the KMeans algorithm instead of 
         spectral clustering and with: $\nclus=6$.}
\label{fig:latent_location_3}
\end{figure}

Finally, Fig.~\ref{fig:latent_location_3} illustrates another clustering method 
based on the KMeans algorithm. It is constructed from the clustering of the INR 
latent vectors and corresponds to $\nclus=6$. Compared to 
Fig.~\ref{fig:latent_location_1}, it illustrates the impact of a different 
clustering method. The KMeans method is a little more naive than spectral 
clustering and makes additional hypotheses about the clusters, in particular 
that they should be spherical. The main difference between the two methods is 
that KMeans produces ``thinner'' clusters. However, the diagonal nature of the 
clusters and the role of magic neutron numbers in identifying clusters remains.

\bibliography{sample}

\end{document}